\documentclass[journal]{IEEEtran}
\usepackage{amsmath,amsfonts}
\usepackage{algorithmic}
\usepackage{algorithm}
\usepackage[utf8]{inputenc}
\usepackage[T1]{fontenc}
\usepackage{array}
\usepackage[caption=false,font=normalsize,labelfont=sf,textfont=sf]{subfig}
\usepackage{textcomp}
\usepackage{stfloats}
\usepackage{url}
\usepackage{verbatim}
\usepackage{graphicx}
\usepackage{cite}
\usepackage{subfig} 
\usepackage{newtxtext} 
\usepackage{newtxmath} 
\usepackage{todonotes}
\usepackage{atbegshi}
\AtBeginDocument{\addtocounter{page}{-1}\AtBeginShipoutNext{\AtBeginShipoutDiscard}}
\begin{document}


\title{Human-like Bots for Tactical Shooters Using Compute-Efficient Sensors}


\author{
    \IEEEauthorblockN{
    Niels Justesen\IEEEauthorrefmark{1}, 
    Maria Kaselimi\IEEEauthorrefmark{1},
    Sam Snodgrass\IEEEauthorrefmark{1}, 
    Miruna Vozaru\IEEEauthorrefmark{1}, 
    Matthew Schlegel\IEEEauthorrefmark{1},
    Jonas Wingren\IEEEauthorrefmark{1},
    Gabriella A. B. Barros\IEEEauthorrefmark{1},
    Tobias Mahlmann\IEEEauthorrefmark{1},
    Shyam Sudhakaran\IEEEauthorrefmark{1},
    Wesley Kerr\IEEEauthorrefmark{2},
    Albert Wang\IEEEauthorrefmark{2},
    Christoffer Holmgård\IEEEauthorrefmark{1},
    Georgios N. Yannakakis\IEEEauthorrefmark{1},
    Sebastian Risi\IEEEauthorrefmark{1},
    Julian Togelius\IEEEauthorrefmark{1}
    }
    \\
    \IEEEauthorblockA{\IEEEauthorrefmark{1}modl.ai}
    \\
    \IEEEauthorblockA{\IEEEauthorrefmark{2}Riot Games}
}
\thanks{\noindent The authors N.J., M.K., S.S., M.V., M.S., J.W., G.B., T.M., S.S., C.H., G.Y., S.R., J.T.(\{niels, maria, sam, miruna, matthew, jonas.wingren, gabriella, tobias, shyam, christoffer, georgios, sebastian, julian\}@modl.ai) are with modl.ai. 

\noindent The authors W.K., A.W. (\{wkerr, alwang\}@riotgames.com) are with Riot Games}




\maketitle

\begin{abstract}

Artificial intelligence (AI) has enabled agents to master complex video games, from first-person shooters like \emph{Counter-Strike} to real-time strategy games such as \emph{StarCraft II} and racing games like \emph{Gran Turismo}. While these achievements are notable, applying these AI methods in commercial video game production remains challenging due to computational constraints. In commercial scenarios, the majority of computational resources are allocated to 3D rendering, leaving limited capacity for AI methods, which often demand high computational power, particularly those relying on pixel-based sensors. Moreover, the gaming industry prioritizes creating human-like behavior in AI agents to enhance player experience, unlike academic models that focus on maximizing game performance. This paper introduces a novel methodology for training neural networks via imitation learning to play a complex, commercial-standard, VALORANT-like 2v2 tactical shooter game, requiring only modest CPU hardware during inference. Our approach leverages an innovative, pixel-free perception architecture using a small set of ray-cast sensors, which capture essential spatial information efficiently. These sensors allow AI to perform competently without the computational overhead of traditional methods. Models are trained to mimic human behavior using supervised learning on human trajectory data, resulting in realistic and engaging AI agents. Human evaluation tests confirm that our AI agents provide human-like gameplay experiences while operating efficiently under computational constraints. This offers a significant advancement in AI model development for tactical shooter games and possibly other genres.

\end{abstract}


\section{Introduction}
\IEEEPARstart{T}{ransitioning} gameplay agents from academia to industry present several challenges. Advanced artificial intelligence (AI) models often require significant computational resources, making them impractical for consumer-grade hardware. In addition, academic models typically aim for optimal performance \cite{silver2018general}, whereas the industry prioritizes human-like behavior to improve the player experience \cite{yannakakis2018artificial}, necessitating a different approach for defining reward functions. There is also the challenge of integrating AI into complex game environments where most computational resources are dedicated to tasks like 3D rendering, leaving limited capacity for AI processes. 

The training process of the AI algorithms often requires processing of large datasets, especially when using pixel-based inputs that contain vast amounts of visual information \cite{pearce2022counter}. In addition, the models themselves are typically large and complex, consisting of many layers and parameters, which require significant computational power and memory to train and run \cite{pearce2022counter, silver2016mastering}. The need for real-time performance in video games further complicates matters, as AI agents must make decisions and react almost instantaneously, a task that is resource-intensive and difficult to achieve with the limited computational budget available in consumer hardware. These challenges highlight the need for more efficient algorithms and innovative approaches to make AI agents practical and effective in commercial video games.

The believability of AI agents is crucial for creating immersive and engaging experiences in multiplayer games, highlighting another key difference between academic research and industry applications. AI agents that can play multiplayer games well and believably are useful in numerous ways \cite{vinyals2017starcraft}. They can act as opponents or teammates when suitable human players are unavailable, serve as stand-ins for specific players who drop out, and even emulate a particular player's style for others to play against. Additionally, they aid game designers by automatically play testing maps, items, and game tweaks, and can be integral to tutorials. In scenarios where human demonstrations are available, it is intuitively preferable for an agent to learn directly from these demonstrations, thereby fostering more human-like gameplay.

Here, we introduce a method for training compute-efficient deep neural networks to play a multiplayer team-based first-person shooter in a human-like fashion via imitation learning. For models to be successfully deployed in the production of modern 3D video games, the whole inference process must be completed in single-digit milliseconds on a CPU thread. Therefore, we train relatively small networks and use simulated sensors rather than pixels as inputs. We evaluate our trained bots for performance, inference time, and believability. Performance evaluation is done through match-ups between bots in the game, as well as by playing human vs bot matches. Inference time is validated by comparing the time required by our models with that of existing models designed for games like CS:GO (Counter-Strike: Global Offensive). Believability is evaluated using Turing-test-like experiment protocol on video recordings.

The contributions of this paper are summarized as follows: 
\paragraph {\textbf{Multi-Sensor Data Integration in Virtual Gameplay  Environments}}  Our models utilize sensors to capture spatial characteristics in gamer's movements within virtual environments without delving into pixel-based processing to optimize bot's movements, ensuring that they navigate virtual environments seamlessly and react appropriately to dynamic changes. The data gathered from these sensors are used for training ML bots, enabling compute-efficient models that can be deployed in commercial video games. 

\paragraph {\textbf{Practicality in AI Bots for Gameplay}} The proposed models provide accurate predictions with minimal computational resources and low latency ensuring reliability and transparency. Prioritizing efficiency in inference allows game industry to deliver timely and dependable insights while reducing the risks of prolonged processing and high resource use. A supervised machine learning approach for temporal human trajectories is adopted to train models to closely mimic human behavior by learning directly from human trajectory data. This leads to realistic agent behaviors.

\paragraph {\textbf{Bot Human-Likeness Multi-faceted Assessment}} In our comprehensive evaluation process, we assess the bots' human-likeness through three key steps. Firstly, we analyze the similarity between distributions of bot behavior and generated data. Next, we examine spatial similarity using heatmaps to visualize bot movements in the virtual environment. Finally, we incorporate human evaluation through a questionnaire, gathering subjective feedback on the perceived beliavability of bot actions. Through this multi-faceted approach, we strive to ensure that our bots exhibit behavior that is both realistic and engaging, enhancing the overall experience.

\section{Background}


In this section we cover related works in imitation learning (Section \ref{sec:IM}) and its challenges when it is tasked to learn to play like a human (Section \ref{sec:limitationIM}) in a multi-player setting (Section \ref{sec:multiplayer}). We end this background section with a discussion on agent believability and human-likeness (Section \ref{sec:believe}).

\subsection{Imitation Learning} \label{sec:IM}

Imitation learning agents are normally trained to perform tasks from human demonstrations by learning a mapping between observations and actions \cite{hussein2017imitation}. The idea of learning by imitation has a longstanding history, but the field has been gaining increased interest in recent years \cite{zheng2022imitation} due to the growing number of available datasets. This surge of attention can be attributed to advancements in computing and sensing technologies, coupled with a growing demand for intelligent applications within video game environments \cite{yannakakis2018artificial}. Imitation learning in video games, however, faces several challenges, which can impact the effectiveness and robustness of the learned policies \cite{hussein2017imitation}. For instance, issues related to model stability and the limited exploration \cite{ladosz2022exploration} present in the demonstrated behaviors of experts are some of the challenges that modern imitation learning algorithms face in that domain. We detail such challenges in the section that comes next (Section \ref{sec:limitationIM}).

\subsection{Learning to Play Like a Human: Core Challenges} \label{sec:limitationIM}

Learning to play a game via imitation learning algorithms is challenging for a number of reasons. First of all, imitation learning algorithms face the risk of forgetting previously learned behaviors \cite{liu2023tail}, especially in large imbalanced datasets or datasets that the possible scenarios are not representative. The adoption of sequential modelling approaches and the enhancement of the architectures with LSTM layers prevents the models from catastrophically forgetting \cite{hochreiter1997long}. A recent example is the work of Pearce and Zhu~\cite{pearce2022counter} that applied LSTM-based behavioral cloning architectures to train playing bots for \emph{Counter-Strike: Global Offensive} (Value, 2012). Recent studies adopt transformer based architectures to handle long-range dependencies in sequences. In \cite{shafiullah2022behavior} for instance authors use a behavior transformer  
to predict multi-modal continuous actions while \cite{dasari2021transformers} employ transformer architectures to one-shot imitation tasks. Similarly in \cite{reed2022generalist} transformer-based sequence models are leveraged as multi-task multi-embodiment policies across a wide range of video game environments, showcasing impressive results in few-shot out-of-distribution task learning. Opposed to the current trend of employing large-scale models for imitating sequential decision-making tasks, like game playing, in this paper we rely on simple yet efficient methods that exploit their rich sensor-based perception to act like human players. Our goal is to make such AI models operational and deployable to an actual commercial-standard game.

Imitation learning tends to exploit demonstrated behaviors, which may limit exploration and hinder the discovery of novel strategies or responses to unforeseen situations yielding limited generalization ability of the trained agent \cite{sreeramdass2023generalized, jiang2022uncertainty}. As a response to this challenge, deep reinforcement learning (deep RL) has led to impressive recent AI breakthroughs with regards to agent generalizability in games such as Atari \cite{mnih2013playing}, Go \cite{silver2016mastering}, StarCraft \cite{vinyals2017starcraft}, Dota \cite{berner2019dota}, and Gran Turismo \cite{wurman2022outracing}. In this work we do not rely on the RL paradigm to learn to play a game better than any other human but instead attempt to design human-like bots in a tactical shooter game using a simple yet generic methodology. While the AI agents we train cannot necessarily transfer to other shooter games their perception mechanism and training methods can. 

\subsection{Learning to Play in Multi-Player Games} \label{sec:multiplayer}

Game states in commercial-standard 3D video games are often large and complex, defined by several variables---such as imagery data \cite{pearce2022counter} and sensory data related to player positions, velocities, health, ammunition, etc. Such a game state space is often associated to a multi-dimensional and complex action space \cite{shih2022conditional}. Arguably, the aforementioned spatio-temporal complexity of video game environments combined with a multi-player setting pose collectively a significant challenge for AI algorithms to effectively explore and learn to play those games well. To address this challenge, a popular approach is to sample multiple actions individually to reduce the complexity of the action space from $N^2$ to $N$ at the cost of losing the conditional dependencies \cite{harmer2018imitation}.

Agents have been trained to play \emph{Quake III Arena} (Activision, 1999) in \emph{Capture the Flag} mode---i.e. where two multiplayer teams compete in capturing the flags of the opposing team\cite{jaderberg2019human}. Here, agents were trained by playing thousands of games, gradually learning successful strategies and competing against humans, even when their reaction times were slowed to match those of humans. To achieve strong agents in StarCraft II, deep RL has been combined with a multi-agent tournament setup to produce agents with diverse strategies \cite{vinyals2017starcraft}. 

In contrary to the studies above, in this paper we focus on simple yet high-performing, compute-efficient and deployable approaches for playing multi-player video games in a human-like way.

\subsection{Believability and Human-like Agents}
\label{sec:believe}


Crafting game playing agents within a multi-player setting, that can effectively emulate human behavior---i.e. in a human-like fashion---adds a non-negligible layer of complexity for imitation learning methods given the rich, diverse and subjective nature of human play \cite{milani2023navigates}.
Various studies in the video games literature have focused on the creation and assessment of playing bots in terms of believability and human-likeness (see \cite{togelius2012assessing, hingston20092k, shaker2013turing} among many). Indicatively, in an early attempt \cite{ortega2013imitating} presented a method for assessing the behavioural similarity of different agents playing Super Mario Bros to humans (or to other agents) whereas  \cite{camilleri2016platformer} introduced a level generation method that maximizes the perceived believability of a Super Mario Bros player (human or not). Recently, Zuniga et al. \cite{zuniga2022humans} and \cite{milani2023navigates} leverage variants of Turing tests for video games and crowdsourced a data set of free-form responses to gain insights into the navigation behaviors that human judges perceive as characteristic of AI vs humans. In this paper we tackle human-likeness in a far more holistic manner than mere navigation. We view human-like assessment as a multifactorial challenge and introduce both spatio-temporal metrics and multifaceted video game Turing-like tests to assess behavioral characteristics of tactical shooter play as a whole. 

\section{The \emph{Lyra:Ascent} Game}

\begin{figure}[!tb]
    \centering
    \subfloat[\textsf{\fontsize{7pt}{7pt}}\label{fig:screenshot}]%
    {\includegraphics[scale=0.12]{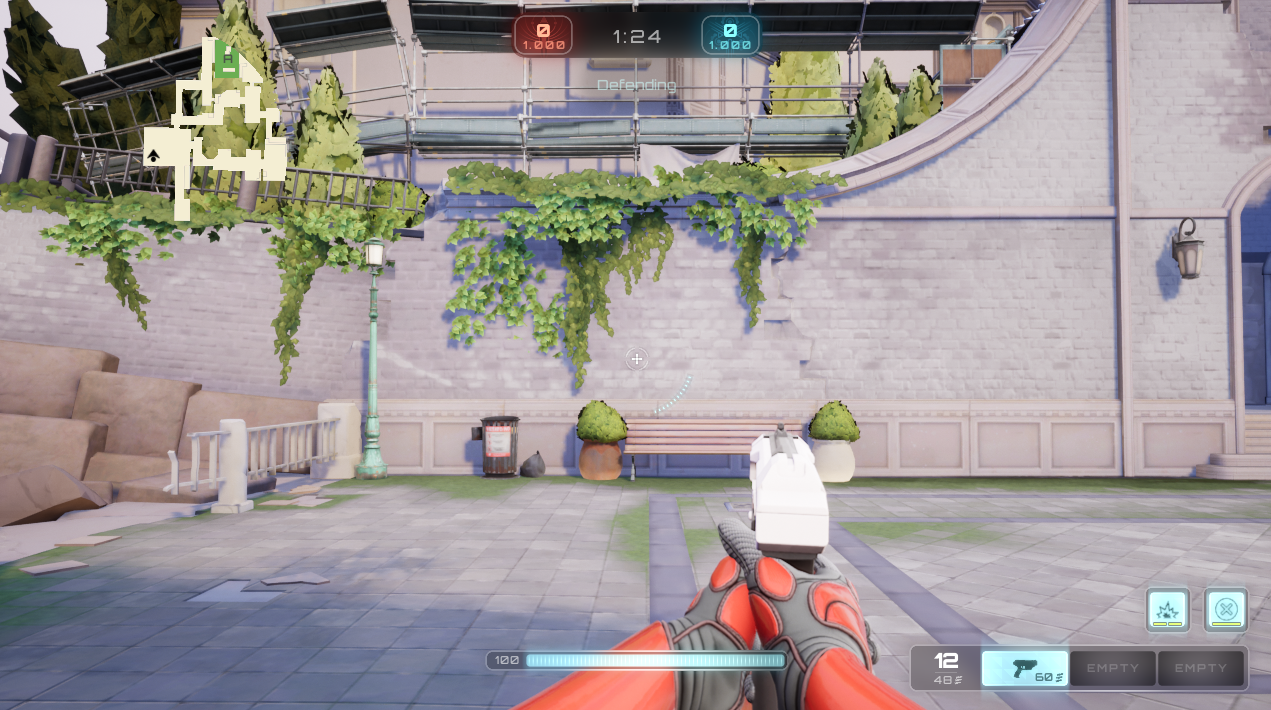}}%
    \hspace{0.5cm} 
    \subfloat[\textsf{\fontsize{7pt}{7pt}}\label{fig:bomb}]%
    {\includegraphics[scale=0.12]{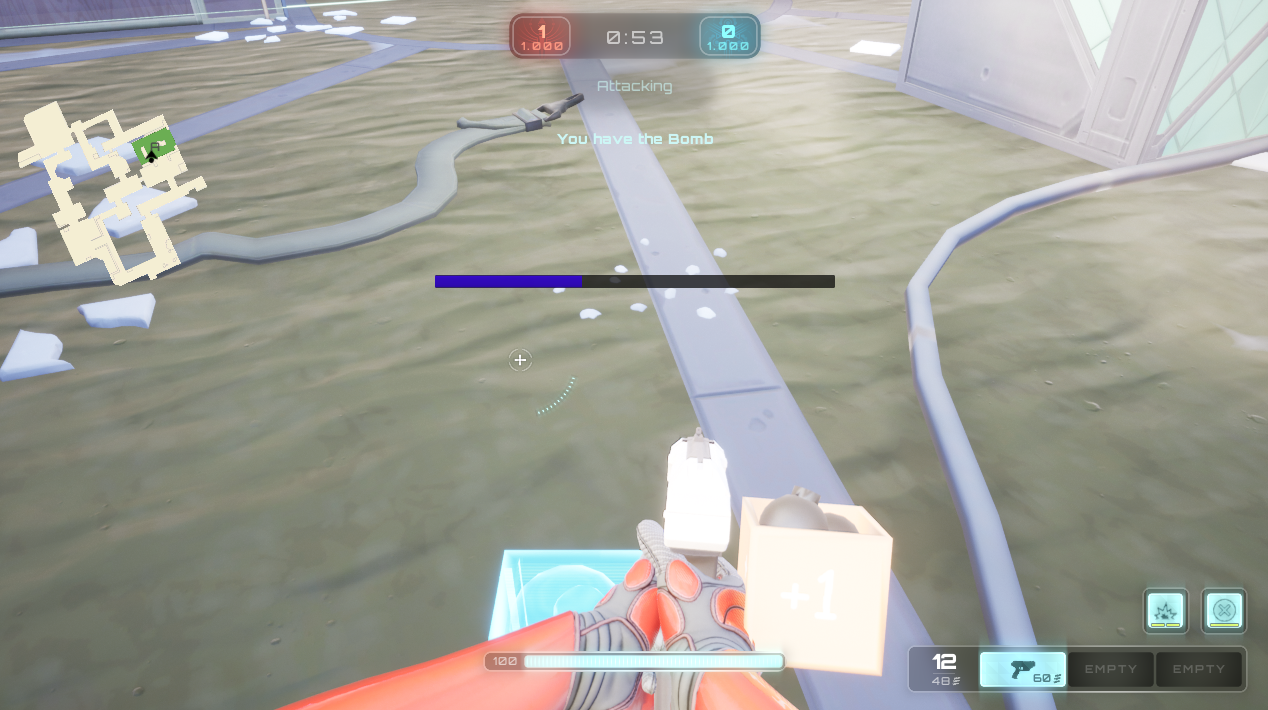}}%
    \hspace{0.5cm} 
    \subfloat[\textsf{\fontsize{7pt}{7pt}}\label{fig:grenade}]%
    {\includegraphics[scale=0.13]{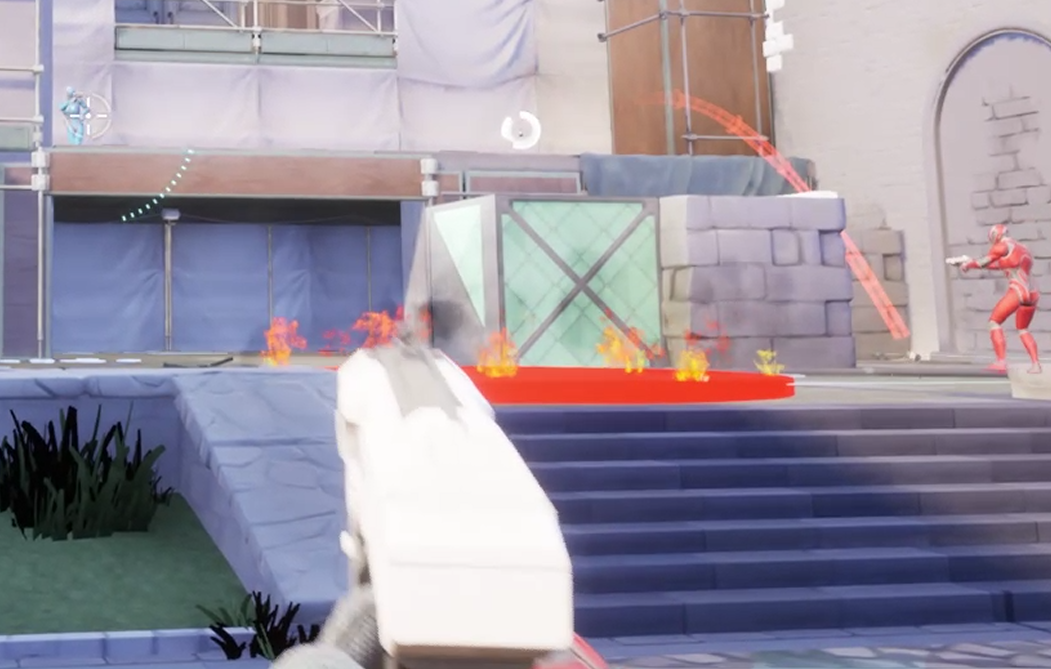}}
    \caption{Selected screenshots of the \emph{Lyra:Ascent} game with (a) defending player at the start point, (b) a player planting the bomb, and (c) an area affected by a grenade. }
    \label{fig:screenshots}
\end{figure}

For this study, we developed a 3D tactical shooter game heavily inspired by \emph{VALORANT} (Riot Games, 2020). We call this game \emph{Lyra:Ascent} and it is built using the Unreal Engine on top of the Lyra framework to ensure our test-bed follows commercial standards. Our game comes with a level similar to \emph{Ascent} in VALORANT which is used for all experiments reported in this paper. Fig.~\ref{fig:screenshot} shows a screenshot of \emph{Lyra:Ascent}. Each match consists of four players paired into two teams: defenders and attackers. The attackers' goal is to successfully plant and detonate a bomb within the bomb site; Fig.~\ref{fig:bomb} shows a player attempting to plant the bomb. The defenders' goal is to defend the bomb site, winning if they successfully defuse a planted bomb or prevent the bomb from being planted. Furthermore, attackers win if all defenders are killed, and defenders win if all attackers are killed before the bomb is planted.

\emph{Lyra:Ascent} is a tactical shooter, and as such its main mechanic is tactical navigation around the map to key locations, and eliminating enemies by shooting or using abilities. Characters in tactical shooter games often come with a large variety of abilities. To focus our study, we opt to adopt only two representative character roles. Specifically, all players on \emph{Lyra:Ascent} are equipped with the same weapon (a pistol), and are given one of two character roles: the \emph{Controller} or the \emph{Initiator}. The \emph{Controller} aims to assist the \emph{Initiator} by decreasing the enemies' access to a target area. He may achieve that either by throwing an incendiary grenade (as seen in Fig.~\ref{fig:grenade}) or by launching a visual blocker in the space. The \emph{Initiator} on the other hand acts more aggressively as it is equipped with skills that may stop enemies from using their own abilities or completely blind them, thereby making them easier to eliminate from the game. 

\begin{figure}[!tb]
    \centering
    \subfloat[\textsf{\fontsize{7pt}{7pt}}\label{fig:map}]%
    {\includegraphics[scale=0.1]{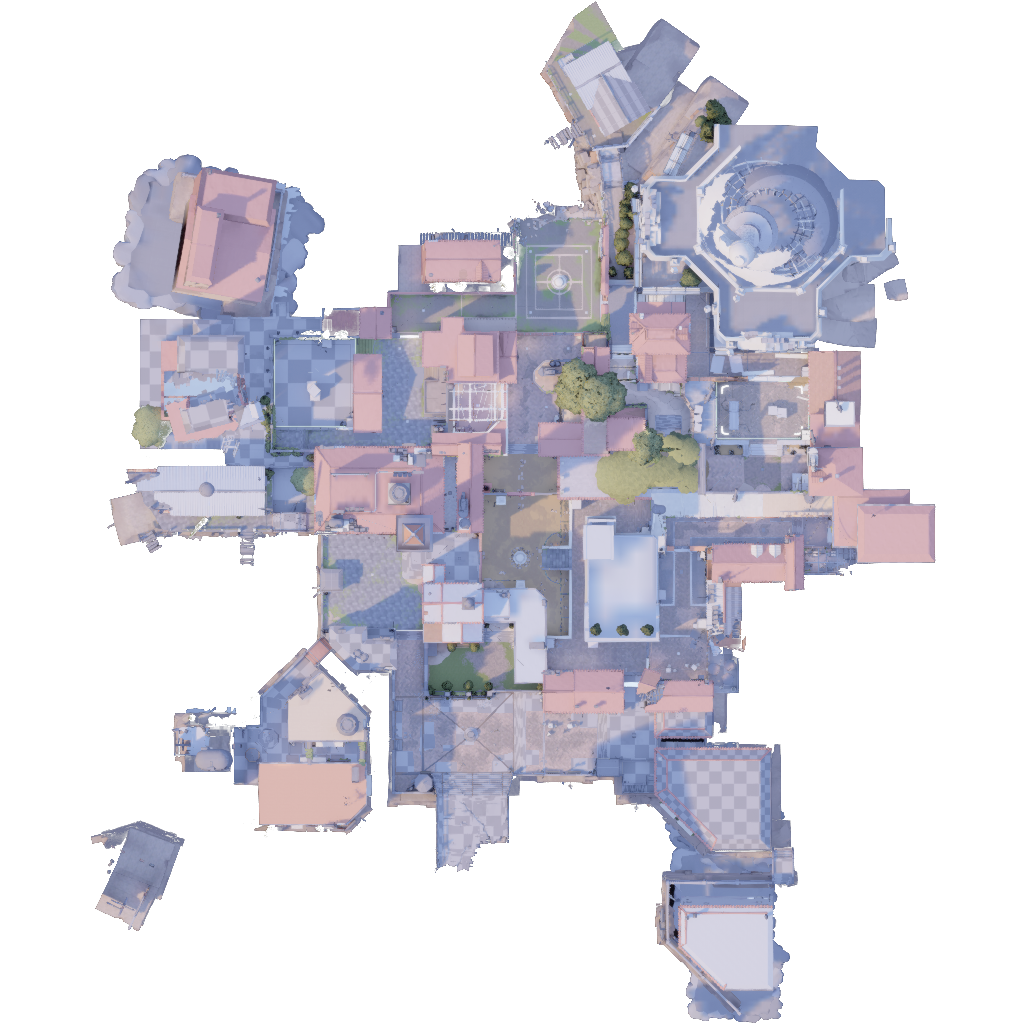}}%
    \hspace{0.5cm} 
    \subfloat[\label{fig:mapschema}]%
    {\includegraphics[scale=0.2]{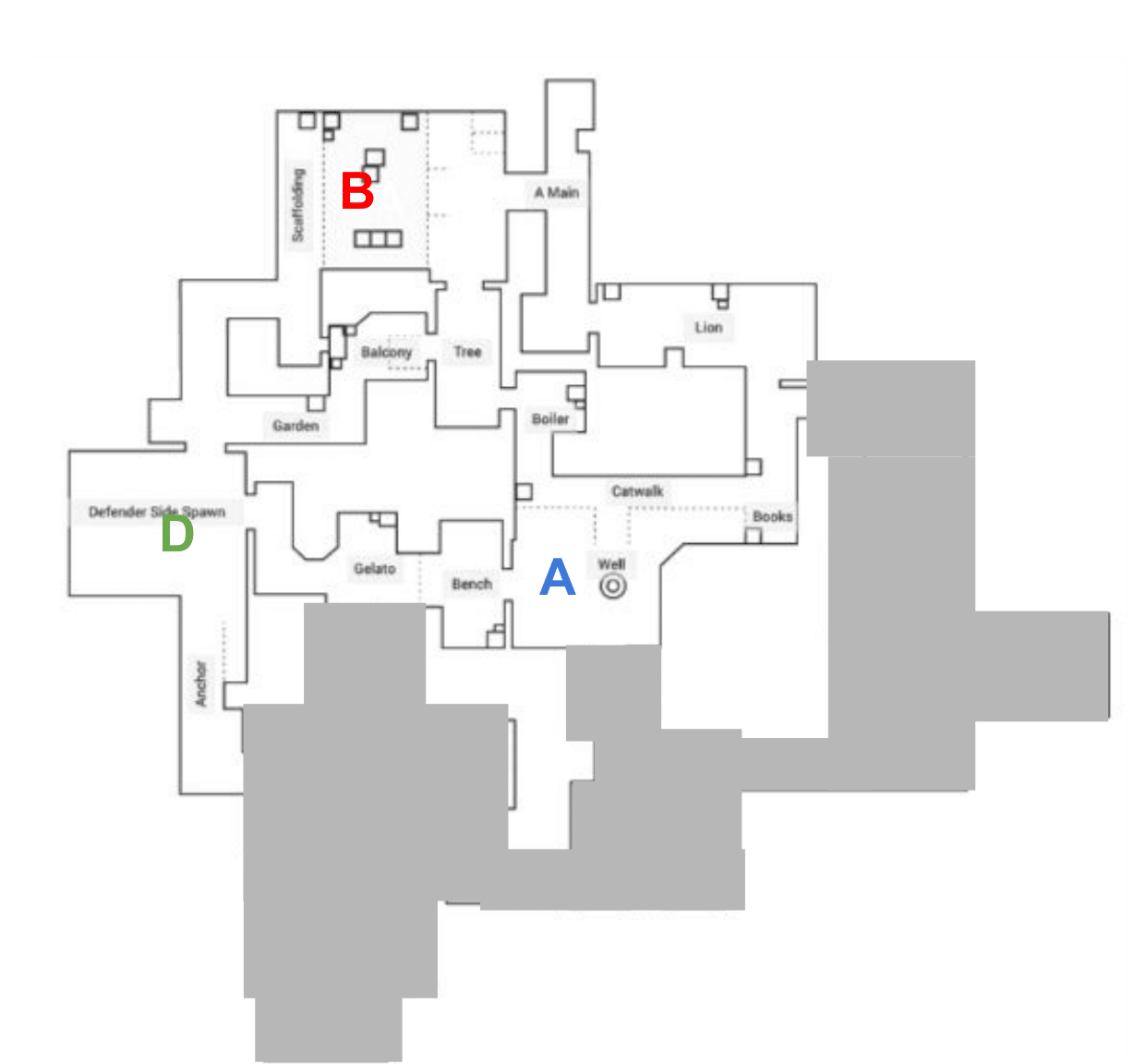}}
    \caption{The \emph{Lyra:Ascent} map used for our experiments. The map is heavily inspired by the Ascent level of \emph{VALORANT} (Riot Games, 2020). (a) Shows a view of the level created in Lyra:Ascent, and (b) shows a map schematic of the level where the {\color{blue}A} represents the attacker spawn point, the {\color{green}D} represents the defender spawn point, and the {\color{red}B} is the bomb site.}
    \label{fig:maps}
\end{figure}

The \emph{Lyra:Ascent} game map we use in this paper consists of a target bomb site and two spawning areas, as seen in Fig.~\ref{fig:maps}. Each team starts in one of the two spawning areas, with the closest one to the bomb site belonging to the defending team.

\section{The \textit{Lyra:Ascent} Observation and Action Space}\label{sec:obsandactions}


In imitation learning, the agent learns by observing and mimicking the behavior in expert demonstrations. This involves defining the observation and action spaces of the environment. The observation space for our agents consists of visual sensors (Section \ref{sec:vis}), audio sensors (Section \ref{sec:aud}), direction sensors (Section \ref{sec:dis}) and game-state information (Section \ref{sec:sta}) which are described in detail in the corresponding sections below. In Section \ref{sec:act} we describe the actions space. 

\subsection{Visual Sensors}
\label{sec:vis}
In a competitive shooter game, it is crucial that game players have precise visual information that, in turn, would allow them to move and aim precisely. We employ range finders that measure the distance between the controlled character and various objects of interest via ray casting. To achieve higher degrees of precision, one usually increases the number and density of range finders. However, range finders are computationally expensive when applied in large numbers and are typically not feasible to achieve high-resolution sensory input for ML models. Range finders that are uniformly distributed in every direction result in an unnecessary amount of detail near the mid- and far-periodic vision areas which will then require larger models for processing such detail.

Inspired by human visual perception, we instead distribute a minimal number of range finders that are dense near the player's crosshair (i.e.\ the center of the screen) and sparse at the far peripheral (i.e. the top, bottom, corners, and both sides of the screen). As a result, a game playing agent may need to adjust the mouse multiple times to accurately locate enemies appearing in its peripheral vision. This behavior mimics the imperfect mouse control of human players in similar occasions. Human players, however, can in contrast to our agents focus their visual perception on away from the crosshair without moving the mouse.

\begin{figure}[!tb]
    \centering
    \subfloat[\textsf{\fontsize{7pt}{7pt}}]%
    {\includegraphics[scale=0.09]{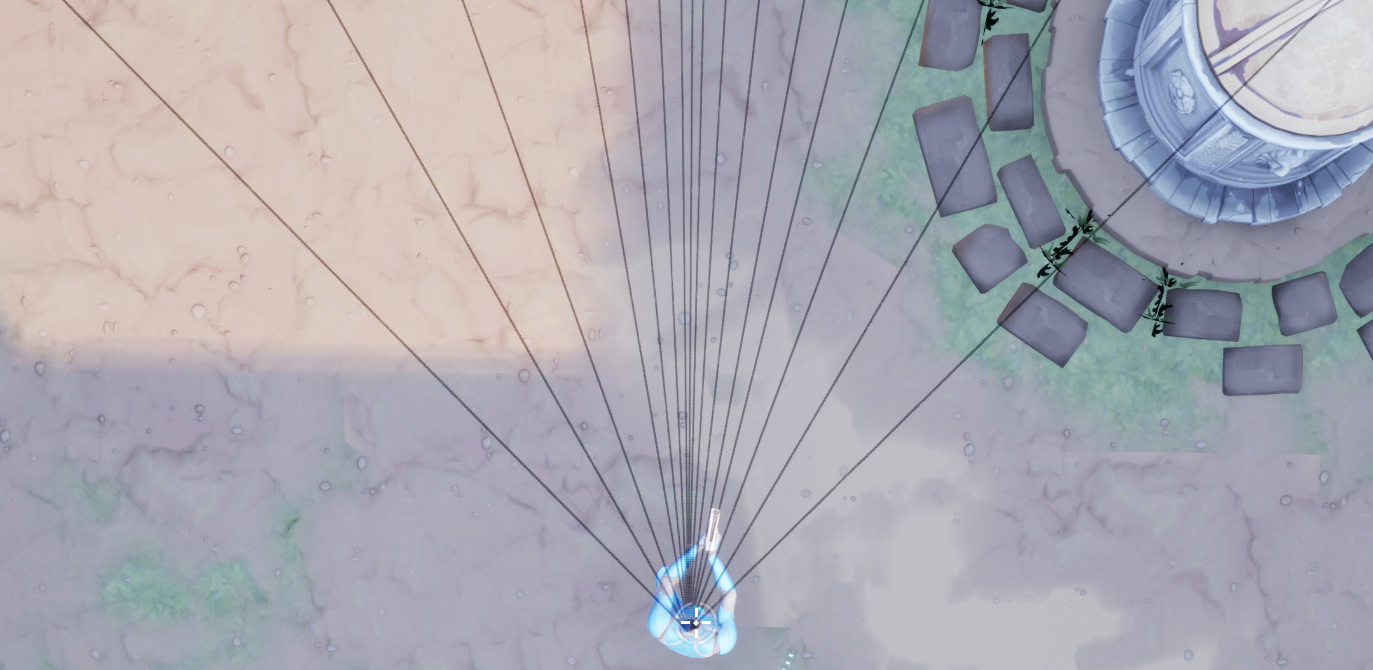}}%
    \hspace{0.5cm} 
    \subfloat[\textsf{\fontsize{7pt}{7pt}}\label{fig:rays-horizontal}]%
    {\includegraphics[scale=0.09]{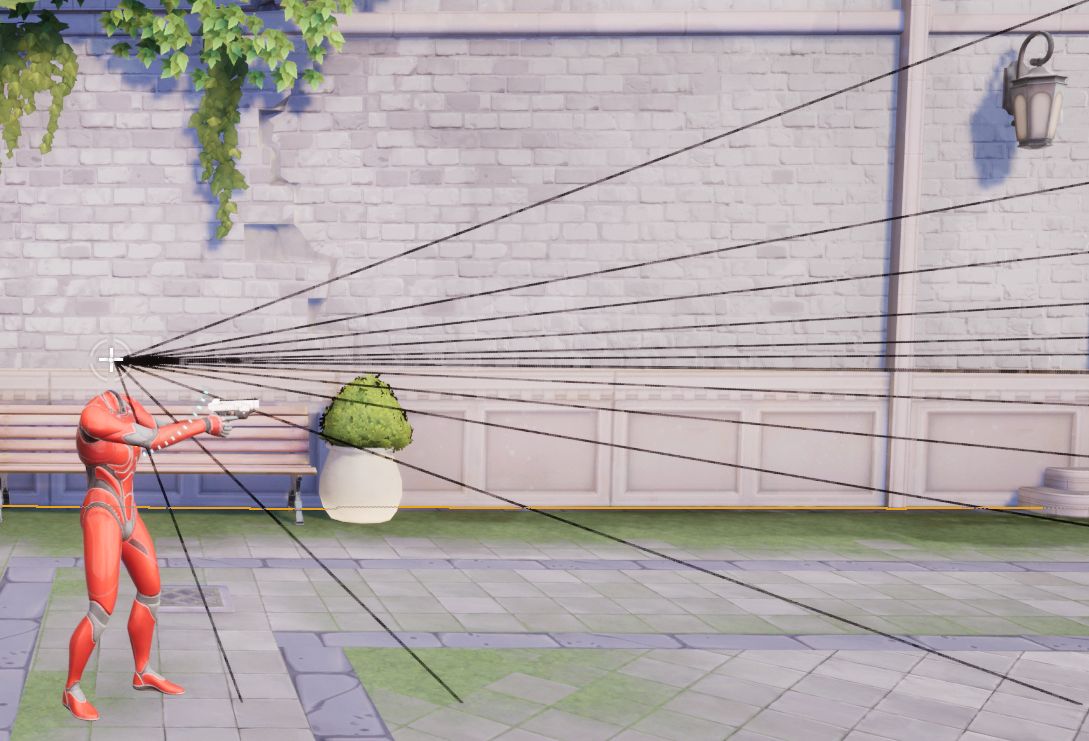}}
    \caption{The figure illustrates (a) 15 horizontal and (b) 15 vertical rays that are distributed non-uniformly, forming a $15 \times 15$ grid-based sensory input.}
    \label{fig:rays-vertical}
\end{figure}

The visual perception system of our bots casts rays across 225 directions, forming a $15 \times 15$ grid for eleven different types of game objects (Fig.~\ref{fig:rays-vertical}). The ten game objects that can be seen are: enemies, teammates, enemy grenades, team grenades, smoke, fire, dropped bombs, planted bombs, bomb site, and finally a layer for all remaining objects such as walls and other static elements in the environment's geometry. Collectively, the visual sensors form a 3D tensor of size $15 \times 15 \times 10$ that is suitable for convolution. Our range finders are casting rays directly from the character's forehead. One of the rays is cast directly forward, while the rest are rotated along the character’s pitch (vertical axis) and yaw (horizontal axis) with the following angles $[70, 45, 20, 10, 6, 3, 1, 0, -1, -3, -6, -10, -20, -45, -70]$ for yaw and $[45, 30, 20, 10, 6, 3, 1, 0, -1, -3, -6, -10, -20, -30, \\ -45]$ for pitch. We include angles that go beyond a human player's vertical vision to ensure that our agent is aware of the terrain and other objects to avoid near its feet. If a range finder hits a game object, its distance to the agent is added to the corresponding visual feature layer. If a ray does not hit an object then the sensor's default value is set to the maximum distance of 100 meters; all visual sensors are finally normalized within this range.

One challenge with our non-linear distribution of range finders, is that important game objects, such as grenades or enemies far away, can be missed in between two rays, especially at the mid or far peripheral areas due to their sparsity. To address this, we cast a small set of additional range finders---one for each important game object in the scene---within the character's 90-degree field of view, which includes all other players, grenades, fire, bombs, and the bomb site. The distance of these additional range-finders are tracked and stored within the nearest cell in the corresponding feature layer.

\subsection{Audio Sensors}
\label{sec:aud}
Audio is an important aspect of shooters as it can provide decisive information such as the location of enemies and when shots are being fired. Such information can help agents make better decisions and exploit silent walking and crouching as a viable tactic during play. Our agents are equipped with 8 evenly distributed directional audio sensors for each one of the following sound types: footsteps and jump sounds performed by other players, shots fired, bomb beeping, grenades exploding, and the bomb being dropped. The audio sensors are based on the sound engine of Unreal Engine 5 (UE5) ensuring that game playing agents can hear and process the same sounds human players can. Each audio sensor stores the distance to each sound type that is less than 100 meters away and is min-max normalized using the maximum range of 100 meters. In contrast to the visual sensors, these are inverted, such that loud/nearby sounds produce a high sensory input value, and sound far away, or no sound, produces low input.

\subsection{Distance and direction}
\label{sec:dis}
The data gathering includes position variables and directions for players, teammates, bomb sites, and bombs. Specifically, X and Y coordinates are combined to determine the positions of various entities. Following this, euclidean distances are calculated between the primary entity's position and the positions of other relevant entities. This process helps in deriving spatial relationships and proximities critical for subsequent analysis and modeling.

\subsection{Game State}
\label{sec:sta}
Game state information refers to additional information about the state of the current match and the team that is not directly collected from visual or audio sensors. Such information includes \emph{player}-related features such as whether the player is jumping or crouching, \emph{bomb}-related features such as whether the player or one of their teammates has the bomb and finally, \emph{team}-related features such as the minimum distance to the nearest enemy and the time left in the current game round. Table \ref{table:gamestate} in Appendix gives details about the game state features in \emph{Lyra:Ascent}.

\subsection{Action Space}
\label{sec:act}
The action space has two parts; one for aiming (mouse movement) and one for all the keys that can be pressed and clicked. 

\begin{figure}
   \hspace*{-0.6cm}
        \includegraphics[width=1.05\linewidth]{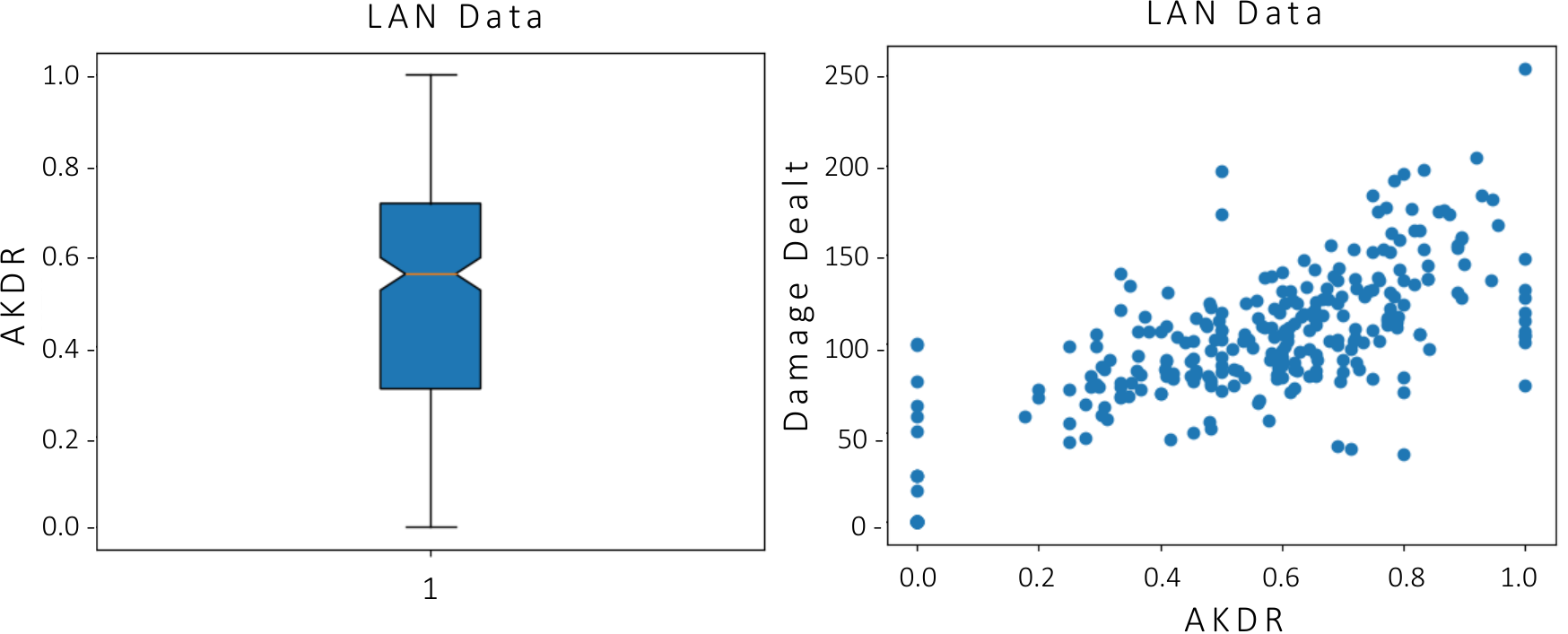}%

    \caption{AKDR vs average damage dealt per round by the player.}
    \label{tmp:datasetstatistics:kdr}
\end{figure}

\begin{figure}
    \hspace*{-0.6cm} \includegraphics[width=1.15\linewidth]{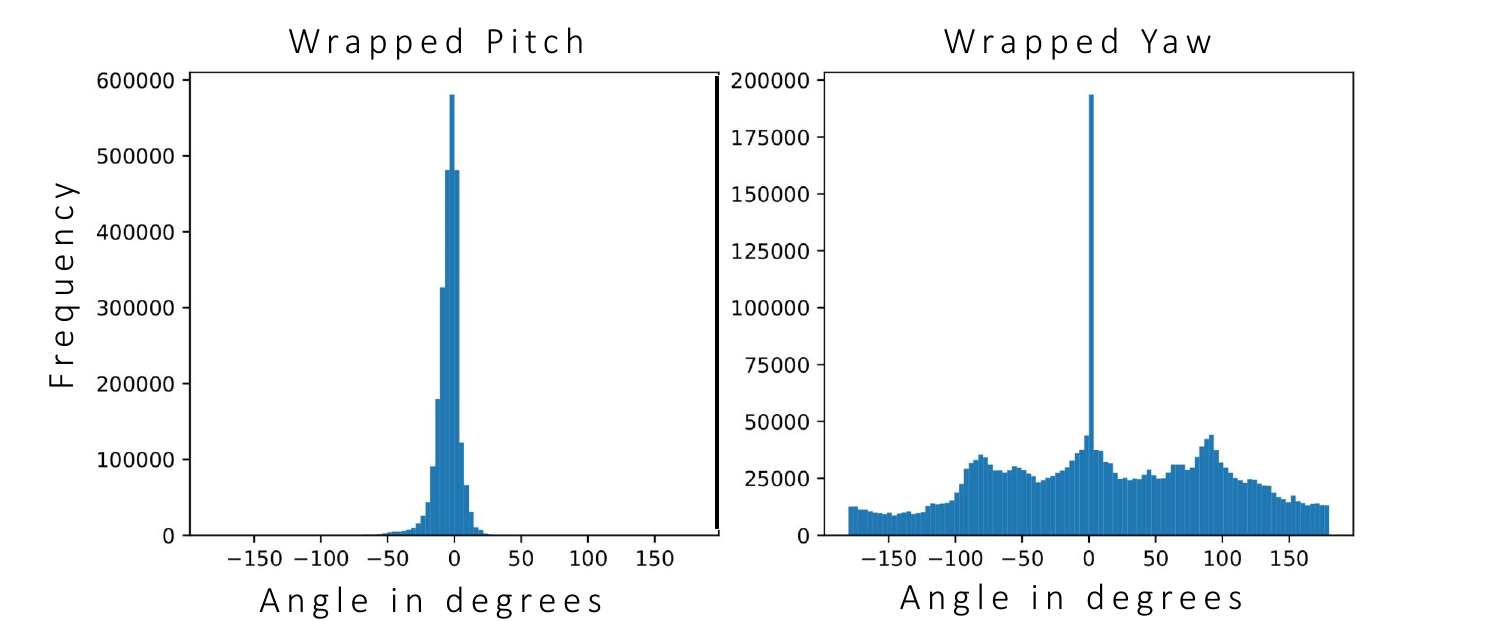}
    \caption{Pitch and Yaw distributions of the players.}
    \label{tmp:datasetstatitiscs:mouse}
\end{figure}

\paragraph{Aiming}Instead of aiming by controlling the mouse, our agent can directly manipulate the rotation of the character's body. The action space for aiming is tightly connected to our visual sensors by offering the same set of actions equal to the angles used by the range finders. The uneven distribution of angles allows for precision when making small adjustments to the aim while also reducing the number of actions at the mid and far peripherals, resulting in a smaller action space. It potentially also makes it easier to learn a mapping between sensors and actions as there for each sensor is an action for that exact direction. 
With this setup, if the agent wants to aim towards something it sees, it has a corresponding action to do so.
After investigating the distribution of actions performed by humans, some of which can be seen in Figure \ref{tmp:datasetstatitiscs:mouse}, we decided to squeeze the top and bottom actions, resulting in the following set of angles for pitch $[20, 10, 6, 3, 1, 0, -1, -3, -6, -10, -20]$. This discrete and unevenly distributed action space is inspired by \cite{pearce2022counter}. In their work, yaw and pitch are sampled independently but we found this approach to produce undesired behaviors when multiple attractive targets are in vision. To solve such issues we combine the yaw and pitch angles to construct a 2D action space with $11\times15=165$ options. Our approach directly couples sensors and actions, in the sense that there exists one action that directly corresponds to each angle used by our range finder sensors. This is a novel approach that to our knowledge has not been done before in this domain. 

\paragraph{Key Presses} The other part of the action space is concerned with key presses and includes: W, A, S, D (for forward, left, backward, right), Space (for jumping), 4 (bomb planting or defusal), G (dropping the bomb), R (reloading), Q (main ability), E (secondary ability), and left mouse click (shooting). The 4-key has to be held down for either four or seven seconds to plant or defuse the bomb. Every other key will trigger an action as soon as they are pressed.





\begin{table}[tbh]
    \centering
        \caption{Presented objectives to players during data collection.}
    \begin{tabular}{p{0.2\linewidth}p{0.3\linewidth}p{0.3\linewidth}}
    \hline
         & \textbf{Attacking} & \textbf{Defending} \\ 
         \hline
         Lava + Smoke & Find the best way to defend the bomb from being defused. & Catch the other team off-guard, be unexpected. \\ 
         \hline 
         Flash + Ability Blocker & Find the best way to plant the bomb. & Make people look at you and not your teammate. Be distracting.\\
         \hline
    \end{tabular}

    \label{tab:prompts}
\end{table}

\section{Dataset}
A dataset was collected from several LAN parties where different players faced off in teams playing several rounds each. We had 28 players produce a total of $48.6$ hours of gameplay. 
Prompts were given to each player according to the class they were assigned which can be seen in Table \ref{tab:prompts}. The teams swapped between attacking and defending on every round but maintained the same role throughout the match. The players had a diverse set of proficiency at playing first-person tactical shooters. The kill-death ratio (KDR) was calculated as $\frac{\text{number of kills}}{\text{number of deaths}}$ approximating how skilled each player is in killing enemy players. The assist-kill-death ratio (AKDR) approximates each player's contribution in killing the enemy team and was calculated as $\frac{\text{kills}+\text{assists}}{\text{kills}+\text{assists} + \text{deaths}}$. Figure \ref{tmp:datasetstatistics:kdr} shows the AKDR, indicating a median assist-kill-death ratio at approximately $0.6$ and a long tail distribution. There were several teams who were never killed, indicated with  AKDR of $1$, with the opposing team's AKDR at $0$. This explains the groupings around $0$ and $1$ in Fig.~\ref{tmp:datasetstatistics:kdr} (right). 
The data was then further cleaned and prepared for training. Each round was split into trajectories for each player with 16 observation-action pairs per second, as in \cite{pearce2022counter}, resulting in a total of $2,797,632$ timesteps. 

\section{Network architectures}\label{sec:network_archs}

\begin{figure*}
\centering
     \includegraphics[width=5.6in]{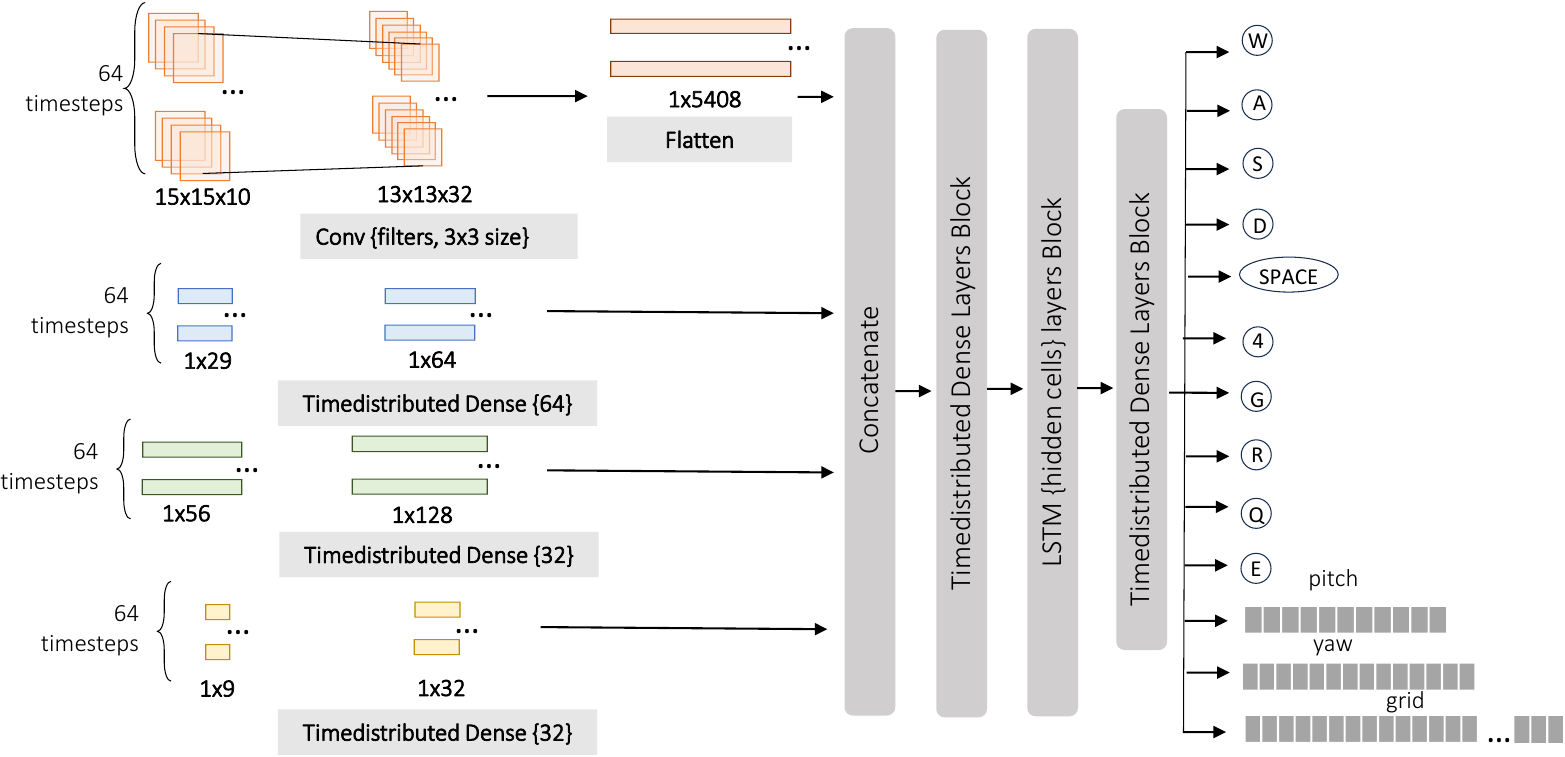}
    \caption{Overview of the architecture and the core blocks utilized. The input features include visual sensors (orange), game state information (blue), audio senors (green), and finally, direction and distance information (yellow).}
    \label{fig:model}
\end{figure*}

Imitating human behavior data exhibits complex dynamics and pose challenges for standard statistical models and machine learning approaches, which typically assume independent and identically distributed (i.i.d.) data. To address this, we adopt a sequential learning approach that incorporates historical information to capture the temporal peculiarities of human behavior trajectory. Memory is a critical component to successfully imitating the human play data, thus, this paper presents findings using memory-based neural networks. Here, we adopt architectures with Long Short-Term Memory (LSTM) units \cite{hochreiter1997long} to enable memory beyond just a few frames. 

We compare various sizes of the LSTM-based architecture in our experimental setup. Fig.~\ref{fig:model} illustrates the general structure of our model. Our problem is a multi-output classification problem with a combination of binary and multi-class outputs. This type of problem is common in imitation learning, where a model learns to mimic the actions of an expert by predicting multiple outputs that represent different aspects of the human's behavior. 

As illustrated in Fig. \ref{fig:model} the encoding layer of the network consists of four parallel streams (with four different colors) for each of the set of features described in Section \ref{sec:obsandactions}. Each layer had an encoding layer associated with the stream. For the standard features, audio sensor features, and distance features this was a simple dense layer of varying sizes with linear activations. A critical observation to the success of our agents was using a convolutional layer for the encoding of the raycast features with dimensions $15\times15\times10$. All the networks used a convolutional layer with kernel size $(3,3)$, stride of $1$, and ReLu activations. Each of the different sized architectures used a different number of learnable filters. After the initial parallel streams, the outputs of these layers are flattened and concatenated to dense layers with ReLu activation. 

After the initial encoding network, the resulting features are passed to an LSTM layer with varying size. Each of the LSTM layers used dropout of $0.5$. After the LSTM layer(s) the state was passed to a set of dense layers using ReLu activations. The final output of the network is then separated into several parallel streams corresponding to the different dimensions of the agent's actions each using an appropriate activation for their loss functions.
Table \ref{tab:network_architecture} shows in detail the selected hyper-parameters of each model as well as its size. 

Following the architecture above, this paper presents networks A, B, C, D, E, and F of varying size. The smallest models A, B, and C are structurally identical besides the number of parameters in each layer. D includes two dense layers instead of one after the LSTM. E and F include two LSTM layers as well as multiple dense layers before and after the LSTM block. Each of these networks are trained with the same procedures (see Table \ref{tab:network_architecture}), and are compared in Section \ref{sec:results}.

\begin{table}
    \centering
     \caption{Size and layout of each model. Each part of the network has its own columns. Comma-separated values indicate multiple layers. \\Additional training parameters: learning rate: $0.0003$, decay: $0.001$, batch size: $96$, epochs: $600$, timesteps: $64$. }
     \resizebox{1.0\linewidth}{!}{%
    \begin{tabular}{cp{1.5cm}p{0.7cm}p{2.0cm}p{1.5cm}p{1.8cm}}
    \hline
         Model & Network Size & Conv. filters &  Dense  block & LSTM block & Dense  block \\
         \hline
         A & $628$, $226$    & $8$  & $256$ & $128$ & $64$ \\
         B & $2$, $348$, $698$  & $16$ & $512$ & $256$ & $128$ \\
         C & $5$, $686$, $298$  & $16$ & $512$ & $768$ & $256$ \\
         D & $14$, $881$, $386$ & $432$ & $1024$ & $1024$ & $512$, $256$ \\
         E & $25$, $373$, $290$ & $32$ & $1024$, $1024$ & $1024$, $1024$ & $1024$, $512$, $256$ \\
         F & $36$, $382$, $490$ & $48$ & $1792$, $1024$, $1024$ & $1024$, $1024$ & $1024$, $512$, $256$\\
         \hline
    \end{tabular}
    }
   
    \label{tab:network_architecture}
\end{table}

\section{Training methods}

Assuming a game environment with states $x \in X$ and actions $a \in A$, the imitation learning problem is to leverage a set of expert demonstrations $U =\{(a_1, x_1),..., (a_N, x_N)\}$ 
to find the probability distribution over possible actions
that imitates the demonstrator's policy as closely as possible.

Behavior Cloning uses a set of demonstrations $(a_i, x_i) \in U_i, \;\forall i \in \{1,..., N\}$ to learn a policy $\pi$ that imitates the state-action mapping in $U$. This can be accomplished through supervised learning techniques, where the difference between the predicted action and the expert's action is minimized with respect to some metric. Concretely, the goal is to solve the optimization problem $ \theta = \underset{\theta}{\arg\min} \sum\limits_{i}^{N} L(a_i, \pi_\theta(\hat{a}_i| x_i))$ where $\pi_{\theta} (\hat{a_i}| x_i)$ is the probability distribution over possible actions for a given state.

To solve this optimization problem we use back propagation when training the models described in Section \ref{sec:network_archs}. 
Additionally, since our model architecture includes LSTM layers we use back propagation through time or BPTT \cite{werbos1990backpropagation}, and specifically the truncated version.
BPTT essentially works by unrolling the recursive parts of the network, and treating each time step as a state and action pair, i.e., the $(a_i, x_i)$ mentioned above. 


\section{Results}\label{sec:results}

To evaluate the practicality and the human-likeness of our trained models, we first investigated the inference time needed by the trained models and then we conducted a two-step human evaluation process. Initially, each model controlled all four players in a 2v2 match, generating four hours of data for both attacking and defending scenarios. For these two scenarios we perform quantitative comparisons between the generated bot data and the human dataset used for training. In particular, we compare the distributions of different behavioral characteristics, including: speed, round length, shots per round, shots per kill, kills per round, and abilities per round. These occurrences are counted for both the attacking and defending sides. As regards the attacking scenario: bomb plant attempts and successful bomb planting attempts, and for the defending side: bomb defusal attempts and successful bomb planting attempts. We note that all the event occurrences are per-player, not per team.

\subsection{Inference Time}
Evaluating CPU inference times allows for informed decisions regarding model selection based on performance requirements and constraints. We compare the inference time of our models to a similar model from the literature that also uses convolutional and LSTM layers that was trained for CS:GO \cite{pearce2022counter}. We use a medium-range desktop gaming PC with a Intel(R) Core(TM) i7-10700KF CPU @ 3.80GHz processor and 32 GB RAM and run inference on the CPU. Table \ref{tab:inference} shows the size of the models we trained and the inference time of those models on the above hardware. All of our models significantly outperform the CS:GO model in terms of inference time. Notice, that our F model has six times as many parameters but still runs significantly faster. Additionally, our C model is comparable in size but is almost five times faster. This is likely due to the high cost of running convolution on large pixel input, where our low-resolution input requires far fewer of these operations. The key insight here is that pixel frames have a uniform distribution of data across the entire frame where our sensors are dense only where precise information is important and sparse where it is less important. This difference is likely smaller when running inference on a GPU. However, in our experience, the overhead of running these relatively small models on the GPU makes it slower than running them on the CPU. 

%


\begin{table}
    \centering
    \caption{CPU Inference time.}
     \resizebox{0.8\linewidth}{!}{%
    \begin{tabular}{cccc}
    \hline
Model & Parameters & Inference Time [ms]\\
\hline
A & 628,226 & $1.95 \pm 0.35$\\
B & 2,348,698 & $2.71 \pm 0.50$\\
C & 5,686,298 & $5.16 \pm 0.91$\\
D & 14,881,386 & $9.59 \pm 0.86$\\
E & 25,373,290 & $16.85 \pm 1.21$\\
F & 36,382,490 & $18.78 \pm 1.47$\\
\hline
CS:GO \cite{pearce2022counter}& 5,964,475 & $24.10 \pm 3.70$\\
\hline
    \end{tabular}
    }
    
    \label{tab:inference}

\end{table}

\subsection{Distributional Similarity}

We collect gameplay data for each model through a series of matches where they play against themselves. In these matches we collect key properties of the gameplay, such as, the round duration, average movement speed, shots fired, shots per kill, number of kills, bomb planting attempts (for attackers), bomb defusal attempts (for defenders), and number of ability activations.
This evaluation dataset consists of four hours of active player data per model for each side (attackers and defenders). We then compare the distributional similarities of these gameplay properties by using the \textit{Jensen-Shannon divergence}. JS divergence is a symmetric way of measuring the similarity between two probability distributions. JS divergence leverages the \textit{Kullback-Liebler divergence} (KL divergence), which is an asymetric measure of distribution similarity. JS divergence works by computing the KL divergence between the provided distributions ($P$ and $Q$ in Equation \ref{eq:js}) and a mixture of those distributions ($M$ in Equation \ref{eq:js}), and then taking the average of those KL divergence values.
Concretely, JS divergence is defined as:
\begin{equation}\label{eq:js}
  \begin{split}
    JS(P,Q)=\frac{1}{2}(KL(P||M) + KL(Q||M)),  \\where \; M(x) = \frac{1}{2}(P(x)+Q(x))
    \end{split}
\end{equation}
\noindent The underlying KL divergence formula is defined as:

\begin{equation}\label{eq:kl}
    KL(P||Q)=\sum_{x \in X} P(x) \log \frac{P(x)}{Q(x)}
\end{equation}


The comparative analysis between human behavior and that of our agents provides valuable insights into the models' representation fidelity and the degree of similarity between the gameplay patterns reproduced by the models and those observed in human behavior. This analysis helps to validate and later refine the models, and also helps us understand the differences between human and model behaviors.

Table \ref{tab:js_div_chars} shows the JS divergence computed across a number of recorded gameplay and behavioral features, and comparing each of our trained models (A-F) against the recorded human player data.
From Table \ref{tab:js_div_chars}, we can see that model D achieves the best JS divergence across more of the dimensions than any of the other trained models.
In particular model D performs most closely to the human data in terms of shots fired per round (``Shots''), Kills, and Bomb Planting Attempts (``Plt. Attempts'') while on the attacking team, and Round Duration (``Duration'') and Kills while on the defending team.
Furthermore, there is only one measure for which model D performs very poorly, and that is for the Average Moving Speed (``Speed'').
Model D is ranked worst and second worst for this feature while on the attacking and defending team, respectively.
For the other measures, where it is neither best nor worst, it tends to land middle of the pack or near the best.

Note that while model D struggles with movement speed, so do most of the other models!
Model B achieves the best JS divergence score by a wide margin for this feature on both attacking and defending teams.
If we look at Figures \ref{fig:at_sp}-\ref{fig:df_sp} in the Appendix, we can see that there is generally not much overlap for the bot and human distributions for this feature across the models.
The models have a tendency to maintain a higher average movement speed throughout the rounds, shifting their distributions to the right. In fact, for model A we can see that there is no overlap between the model and human distributions!

Another observation of note is that the larger models (E and F), struggle with replicating the Shots Fired distribution from the human data.
E struggles more on attack, and F struggles more on defence.
If we look at Figure \ref{fig:at_sh}, we can see that E's Shots fired distribution is skewed more heavily to the lower values, and we can see the same in Figure \ref{fig:df_sh} for F.
In fact, as the model size grows there is a general trend of the models being more conservative in their number of shots fired, resulting in tail end of the distributions (around the $40-60$ buckets) being more and more sparse.

In Appendix \ref{sec:plots}, we include plots showing the distributions for some of the features in Table \ref{tab:js_div_chars}.
Namely, we have plots for Round Duration (Figures \ref{fig:at_tim}-\ref{fig:df_tim}), Average Movement Speed (Figures \ref{fig:at_sp}-\ref{fig:df_sp}), and Shots Fired per Round (Figures \ref{fig:at_sh}-\ref{fig:df_sh}).
These plots show the human distribution compared to trained bot distributions while on attacking and defending teams.

\begin{table}[t]
    \centering
    \caption{Jensen-Shannon Divergence.}
     \resizebox{1.0\linewidth}{!}{%
    \begin{tabular}{ccccccc}
    \hline
Models &A&B&C&D&E&F\\
\hline
&&& ATTACK&&&\\
Duration&$0.119$&$0.043$&$0.048$&$0.029$&$\mathbf{0.027}$&$0.041$\\
Speed&$0.654$&$\mathbf{0.050}$&$0.148$&$0.545$&$0.403$&$0.303$\\
Shots&$0.010$&$0.016$&$0.010$&$\mathbf{0.008}$&$0.024$&$0.012$\\
Shots/Kills&$0.020$&$0.007$&$\mathbf{0.006}$&$0.010$&$0.015$&$0.011$\\
Kills &$0.011$&$0.004$&$0.004$&$\mathbf{0.002}$&$0.003$&$0.002$\\
Plt. Attempts &$0.003$&$0.004$&$0.003$&$\mathbf{0.001}$&$0.004$&$0.021$\\
Abilities &$0.050$&$0.004$&$0.004$&$0.013$&$0.011$&$\mathbf{0.003}$\\
\hline
&&&DEFENCE&&&\\
Duration&$0.113$&$0.036$&$0.037$&$\mathbf{0.023}$&$0.023$&$0.035$\\
Speed&$-$&$\mathbf{0.249}$&$0.511$&$0.616$&$0.558$&$0.276$\\
Shots&$0.016$&$0.016$&$\mathbf{0.005}$&$0.011$&$0.014$&$0.021$\\
Shots/kills&$0.021$&$0.015$&$0.009$&$0.008$&$\mathbf{0.006}$&$0.012$\\
Kills &$0.006$&$0.001$&$0.001$&$\mathbf{0.001}$&$0.001$&$0.003$\\
Def. Attempts &$0.027$&$0.028$&$0.024$&$0.018$&$\mathbf{0.013}$&$0.016$\\
Abilities&$0.057$&$0.014$&$0.009$&$0.014$&$0.011$&$\mathbf{0.008}$\\
\hline
    \end{tabular}
    }
    
    \label{tab:js_div_chars}
\end{table}

\begin{figure} [!tb]
    \centering
    \includegraphics[width=0.8\columnwidth]{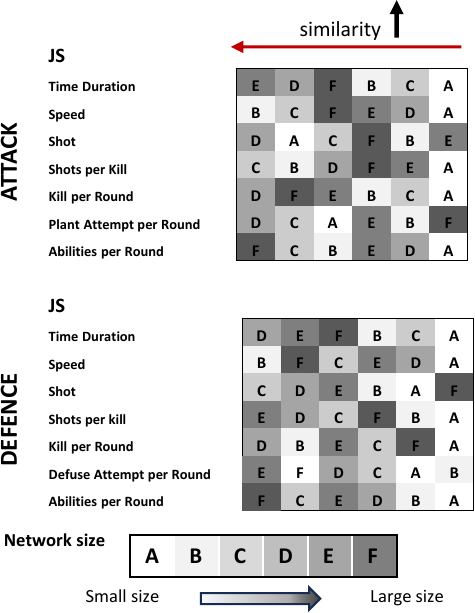}
    \caption{Comparative results based on Jensen-Shannon divergence between the different models. Dark colors illustrate large size models, whereas lighter colors have been selected for the smaller size models. }
    \label{fig:sim}
\end{figure}

Fig.~\ref{fig:sim} shows a visual ranking of the trained models based on their JS divergence scores for each of the compared features. Models located to the left are more similar to the human data compared to those located to the right for each feature. The top grid refers shows the similarities while on the attacking team, and the bottom grid is while on the defending team. From these similarity comparisons we can conclude that model D is the best (quantitatively) performing model from those we trained. Furthermore, we observed that model D is more reactive to enemies and sustains longer fights. 
Model D also follows the enemies, evades shots, and uses abilities in a more targeted fashion during fights.
But these are of course just our anecdotal observations; in Section \ref{sec:human_eval} we walkthrough a study we conducted to determine how realistic and believable others would find model D.




\subsection{Evaluating Spatial Similarity in Human-Like Behavior}
\label{sec:heatmap_analysis}

\begin{table}[H]
    \centering
    \caption{Distance measures comparing the heatmap generated from human training data to a heatmap created for each model A-F. Best value in bold.} 
    \resizebox{1.0\linewidth}{!}{%
    \begin{tabular}{ccccccc}
    \hline
    Models & A & B & C & D & E & F \\
    \hline
    &&& ATTACK &&& \\
    EMD 1D (no location) & 4.4 & \textbf{0.6} & 1.8 & 1.4 & 4.4 & 0.8 \\
    EMD 2D Euclidian & 4.146 & \textbf{1.825} & 2.491 & 2.529 & 3.680 & 2.535 \\
    ASD & 0.603 & 0.449 & 0.407 & \textbf{0.364} & 0.431 & 0.414 \\
    \hline
    &&& DEFENCE &&& \\
    EMD 1D (no location) & 8.6 & 6.2 & 2.6 & \textbf{2} & 3.2 & 2.2 \\
    EMD 2D Euclidian & 4.143 & 4.779 & 3.155 & \textbf{2.689} & 3.681 & 3.844 \\
    ASD & 0.634 & 0.531 & 0.427 & \textbf{0.381} & 0.450 & 0.450 \\
    \hline
    \end{tabular}%
    }
    \label{tab:heatmap_distances}
\end{table}

\begin{figure*}[tbh]
\captionsetup[subfloat]{labelformat=empty}
\centering
\begin{tabular}{@{}ccccc@{}}
\subfloat[Human]{\includegraphics[width = 0.9in]{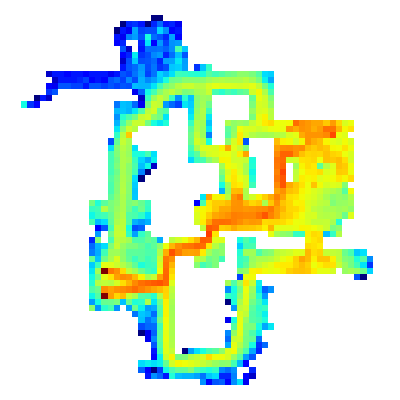}} &
\subfloat[A]{\includegraphics[width = 0.9in]{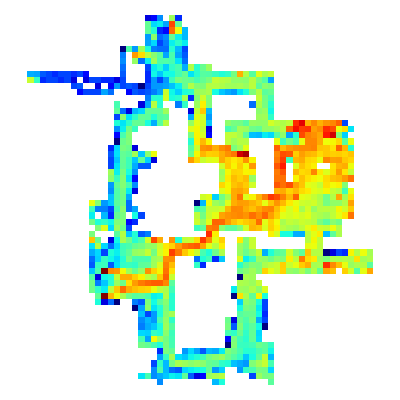}}
\subfloat[C]{\includegraphics[width = 0.9in]{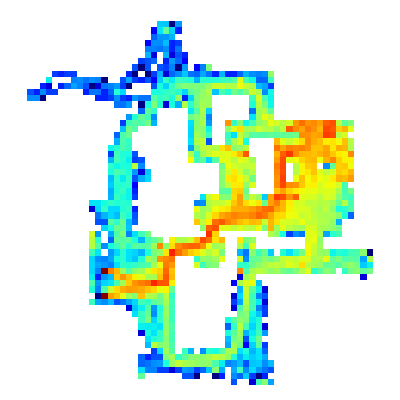}} &
\subfloat[D]{\includegraphics[width = 0.9in]{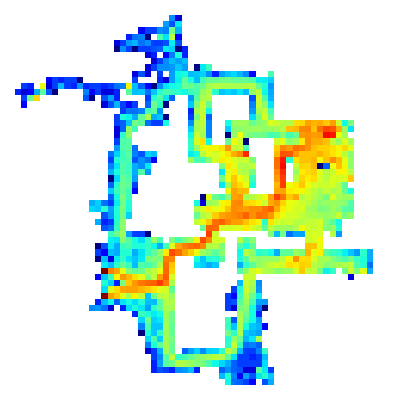}} &
\subfloat[E]{\includegraphics[width = 0.9in]{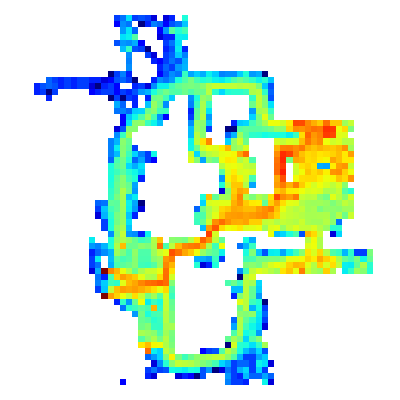}} &
 \\
\subfloat[Human]{\includegraphics[width = 0.9in]{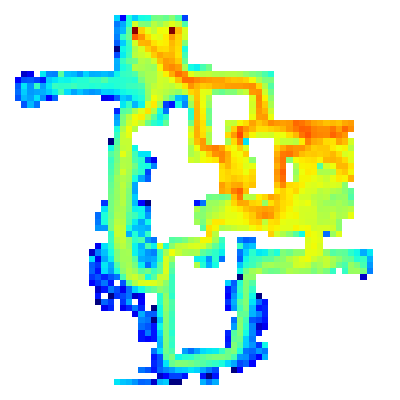}} &
\subfloat[A]{\includegraphics[width = 0.9in]{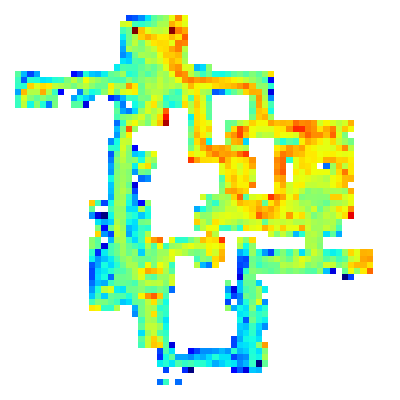}}
\subfloat[C]{\includegraphics[width = 0.9in]{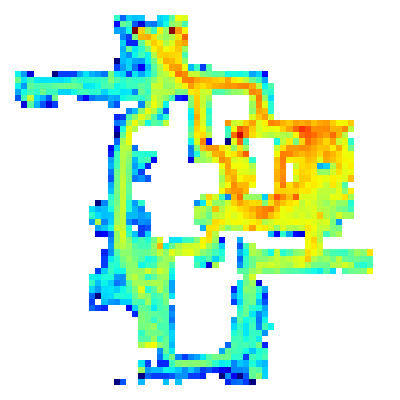}} &
\subfloat[D]{\includegraphics[width = 0.9in]{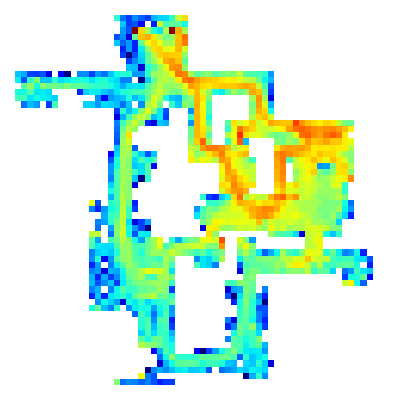}} &
\subfloat[E]{\includegraphics[width = 0.9in]{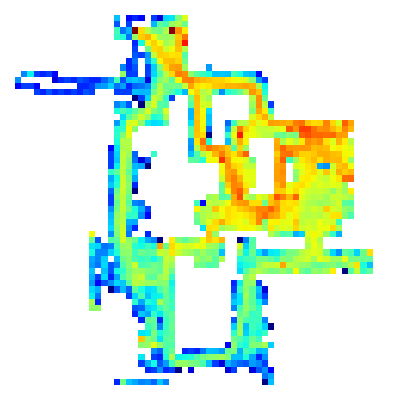}} &\\
\end{tabular}
\caption{Indicative heatmaps showing the map coverage of humans and bots when attacking (top) and defending (bottom). It is noticeable that the smallest model (A) gets stuck in various spots across the map.}
\label{fig:heatmaps}
\end{figure*}

In order to be able to compare the moving patterns of the bots with that typical for human players, we have conducted further analysis using heatmaps imposed on the game map. Heatmaps in the context of an agent playing a video game visualize the spatial distribution of the agent's  presence across different locations on the game map. Each location on the map is represented by a point, with the different points to draw the trajectory of the agent and the colored scale to present the intensity of the agent's presence in each point. Cooler colors such as blue or green indicate areas where the player spends less time or engages in fewer actions. Warmer colors such as yellow, orange, or red indicate areas with high player activity or presence. These colorings can be seen in Figure ~\ref{fig:heatmaps}.

Table \ref{tab:heatmap_distances} contains details about the quantitative comparison of the heatmaps visualising player location for bots and humans. As the heatmaps and the previous section on behavioural characteristics show, different aspects of human behaviour are reproduced by the various models to varying degrees. In the following, we quantify some of these differences in moving patterns by comparing the resulting heatmaps numerically. Since the heatmaps can be considered distributions of how much time was spent in which location, and inspired by previous research \cite{pearce2022counter}, we use the Earth Mover Distance as a way to compare (see also equation \ref{eq:emd}). Earth Mover Distance (EMD) (also called Wasserstein Distance) works by quantifying the minimal work required to transform one distribution into another, and offers a robust metric for comparing human and bot distributions. Let $\mathcal{J} (P, Q)$ denote the set of all joint distributions $J$ for $(x, y)$ that have marginals P and Q. The $J(x,y)$ indicates how much ``mass'' must be transported from $x$ to $y$ in order to transform the distributions $P$ into the distribution $Q$. The EMD then is the ``cost'' of the optimal transport plan and is computed by:
\begin{equation}
    EMD(P,Q)=\inf_{J \in \mathcal{J} (P,Q)} \mathbb{E}_{(x,y)\sim J } [\; ||x-y|| \;] 
\label{eq:emd}
\end{equation}
\noindent EMD can be applied to 2D distributions, thus making it useful for comparing the heatmaps in an interpretable manner.

In order to add context and baselines, we compute EMD between the distributions of how much time was spent in a given cell, ignoring its locations. Each distribution is calculated as a histogram with buckets. This approach thus should be able to identify if one heatmap contains more instances of the player being "stuck" or "camping" than the other. It would, however, not be able to differentiate between the two. Furthermore, we compute EMD on the 2D grid to compare how much time was spent in which location in the map, discretised as cells. We are using Euclidian distance on the grid to determine the cost of moving between the different cells. For example, moving from cell $(0,0)$ to $(4,0)$ incurs a cost of $4$, and so does moving to $(0,4)$ instead. This approach thus compares if the same relative amount of time was spent in the same locations. This reduces the risk of counting a player being "stuck" as intentional "camping", as camping only makes sense in specific locations. Absolute Summed Difference (ASD) of matrix cells as a baseline was also computed.

Fig. \ref{fig:heatmaps} visualises how much time the player and agents spent in each location of the map. From the figure, we can see that the bots' heatmaps more closely mirror the human heatmaps while attacking than while defending; this is further reflected in Table \ref{tab:heatmap_distances} by the generally lower EMD values for attacking over defending. When comparing model performance overall, it seems like there is a sweet spot in terms of model size right around where model D is. The smallest model, A, has some of the biggest distance values observed; and while the larger models, E and F, perform decently overall, they still perform considerably worse than C and D in defense mode. The only exception is the smaller model, B, which performs very well in attacking mode, but really poorly when playing defense. Based on a visual comparison of the heatmaps, the distances obtained by B in attack mode are most likely due to more varied behaviour at the opposing team's spawn point and less frequent utilisation of the balcony area. The latter, however, highly depends on the defending agents' behaviour as this is frequently the location of fights due to proximity to the bomb site. Conversely, when model B is on the defending team, it does not behave like human players. This is reflected in the obtained distance measures and can also be seen in the heatmaps. Namely, the model B bots seem to cover the map much more evenly than human players, especially in the bottom area. Based on the quantitative comparisons of the heatmaps, model D is the most robust in replicating human-like moving patterns across game modes and different measures.

\section{Human Perception of Gameplay Authenticity}\label{sec:human_eval}

This section presents the approach we followed to evaluate our methods via human feedback. We first detail the protocol we used for the purposes of comparing our trained bots against human players (Section \ref{sec:exp_protocol}), we then go through the analysis of the data obtained and we discuss the core findings of this set of experiments (Section \ref{sec:evaluation_data}).

\subsection{Experimental Protocol} \label{sec:exp_protocol}

In this study, we designed a post-game assessment using predefined videos extracted from $2v2$ matches of \textit{Lyra: Ascent} featuring either human players or our trained bots. After the matches, we carefully selected and edited videos showcasing various in-game situations for use in a comparative questionnaire. The purpose of the questionnaire was to evaluate the human-likeness of the bots by asking participants to classify each video as either human or bot, providing valuable insights into the effectiveness of the bots in mimicking human behavior in gaming scenarios.

Each video was $15$ seconds long and was drawn from gameplay footage involving either two human players competing against two other human players or two bots competing against two other bots. We ensured that no videos included periods of inactivity, such as when a player was dead or during transitions between rounds. This selection process resulted in $16$ videos featuring bots and $16$ featuring human players, for a total of $32$ videos. The videos were presented in a randomized order to $20$ participants, all of whom were experienced gamers. Participants could replay the videos as needed to improve the accuracy of their assessments. The protocol was carefully structured to ensure variability and eliminate bias, creating a robust framework for evaluating participants' ability to distinguish between human and bot behaviors under controlled conditions.

Prior to participating in the study, all participants were provided with a detailed information sheet outlining the purpose, procedures, and potential risks associated with the experiment. This document ensured that participants were fully aware of their rights and the nature of the tasks they would be engaging in. Participants were informed that their participation was voluntary and that they could withdraw at any time without any consequences.

\subsection{Data Analysis and Findings} \label{sec:evaluation_data}

\begin{figure} [!tb]
    \centering
    \hspace*{0.0cm} \includegraphics[width=0.55\textwidth]{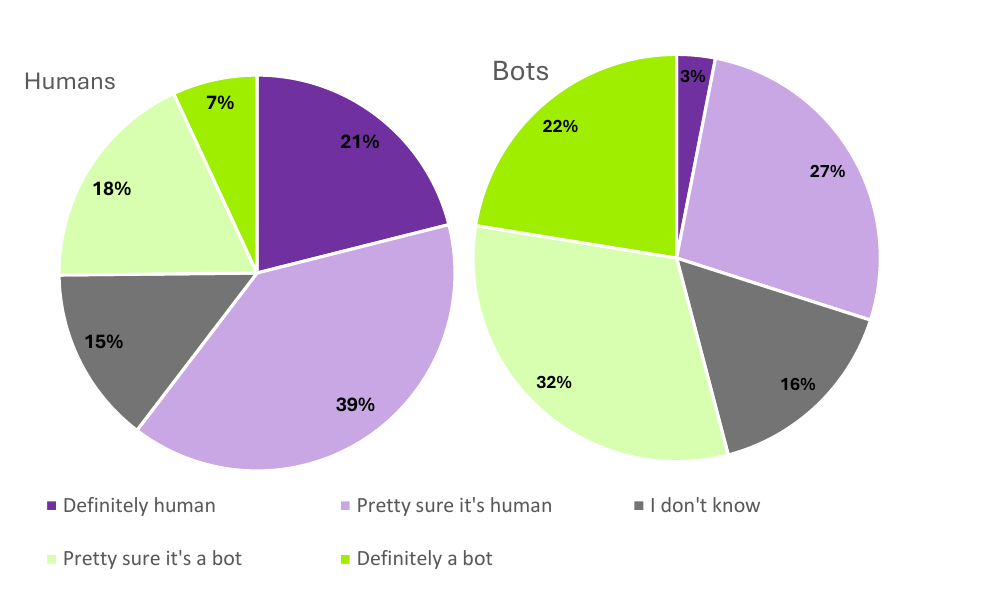}
    \caption{Replies of the human evaluation questionnaire with $32$ questions on whether the presented video clip is human or a bot player. The study involved $20$ participants who evaluated the realism of the bots. }
    \label{fig:eval}
\end{figure}

The human evaluation results provide promising insights into the realism of the bots, demonstrating that they are effective at mimicking human behavior and appearance. When participants were asked to classify videos, $39\% + 21\% = 60\%$ correctly identified the human videos, while $18\% + 7\% = 25\%$ mistakenly labeled them as bots, and $15\%$ expressed uncertainty. The above average identification of humans in the videos suggests that the majority of participants can accurately identify behavioral markers associated with human behavior in the game. This ability of correctly identifying the behavior of humans underscores the effectiveness of the bots in appearing realistic. Notably, $27\% + 3\%=30\%$ of participants were convinced that bot videos were actually human, which highlights the bots' ability to display traits typically associated with human behavior convincingly. While $32\% + 22\%=54\%$ of participants correctly identified the bots, the fact that nearly a third mistook them for humans points to a level of sophistication in the bots' behavior that can deceive viewers. Additionally, only $16\%$ of participants were unsure about the nature of the bot videos, which is comparable to the $15\%$ uncertainty rate for human videos, further emphasizing that the bots are achieving a similar level of ambiguity as genuine human videos and that the behaviors presented were not imbalanced in terms of clarity or other extraneous factors for either humans or bots. 


\section{Conclusion}
This paper introduces a practical approach for creating efficient ML-based bots in an attempt to bridge the gap between academia and game industry. Our proposed imitation learning approach, utilizes deep neural networks with sequential capabilities together with compute-efficient sensors to successfully play a multiplayer team-based first-person shooter game. Our approach is deployed taking into consideration the actual computational constraints arising from real-life game applications and actual game environments, merely focusing in the aspects referred to below:

\paragraph{Data collection} As first step, we collected a dataset of human gameplay trajectories by recording the player's actions and also by setting up a novel array of sensors to capture the spatial and audio related characteristics of the surrounding virtual environment space. Notably, the emphasis is placed on the judicious allocation of computational resources, ensuring optimal performance with minimal latency, thereby fortifying reliability within the gaming ecosystem in near-real-time application scenarios.

\paragraph{Practical human-like ML models} Supervised machine learning leveraging temporal human trajectories is adopted to imbue our bots with behaviors mirroring human-like patterns. This deliberate strategy engenders a sense of realism and immersion, enhancing the overall gaming experience. Our novel sensor system, is an important step towards computational feasibility of neural network policies in commercial video game production. Our best model has almost $15$ million parameters and has an inference time of $9.59$ ms. on average per decision on a consumer-graded gaming PC.

\paragraph{Multi-faceted evaluation of the bot believability} Moreover, our research includes a multi-faceted assessment of the trained bots' human-likeness. Firstly, we analyze the similarity between distributions of bot behavior and collected human data. Subsequently, we check the movement patterns of the bots and compare against the human patterns using heatmaps for visualization and applying similarity measures to compare the time spent in the different areas of the game environment. Finally, we include human feedback through structured questionnaires to evaluate the believability of the bots' behavior.

Based on the analysis and quantitative comparisons, a model (known as Model D in this paper) with roughly $14.9$ million parameters, using convolutional and LSTM layers, is the most robust in replicating human-like behavior, and respecting the trade-off between computational-efficiency, performance, and human-likeness.

\subsection{Future work}
Human behavior is characterized by stochasticity and multi-modality and exhibits structured correlations across action dimensions. 
Advancements in imitation learning algorithms beyond behavior cloning have witnessed a notable convergence with techniques such as reinforcement learning (RL) or behavioral transformers, fostering more adaptive and robust learning frameworks to compensate with human behavior in unseen scenarios. Applying RL to our dataset is an interesting area of research as it has the potential to create diverse behaviors by actively exploring the environment and recovering the optimal reward and agent policy. Using algorithms such as Generative Adversarial Imitation Learning (GAIL) to improve the robustness and diversity of learned policies \cite{ho2016generative} can lead to more robust and naturalistic policies. Recently, approaches such as the Decision Transformer (DT) \cite{chen2021decision} and the Trajectory Transformer (TT) \cite{janner2021offline} were introduced to serve as alternatives for Offline RL algorithms. The Generalized Decision Transformer (GDT), is another variant of the extended transformer models family that can be applied for our dataset\cite{furuta2021generalized, sudhakaran2023skill}.

We view our current results as an initial step toward bridging the gap between academic advancements in human-like gameplaying agents and the game industry's demand for robust and adaptive player bots. We hope this work serves as a foundation and inspiration for further exploration in this domain.

\bibliographystyle{IEEEbib}
\bibliography{main}

\newpage

\appendix
\section{Additional Game State Features}
\label{apx:appendix}

\begin{table}[H] 
\centering
\label{table:gamestate}
\begin{tabular}{ll}
\textbf{\textit{Feature}} &\textbf{\textit{Range}}\\
\hline
\hline

\textbf{Player} &  \\
\hline
Is on attacking team	&	[0;1]	\\
Is jumping	&	[0;1]	\\
Is falling	&	[0;1]	\\
Is shooting	&	[0;1]	\\
Is being shot	&	[0;1]	\\
Is crouching	&	[0;1]	\\
Has main ability Zero	&	[0;1]	\\
Has main ability Sky Smoke	&	[0;1]	\\
Has secondary ability Incendiary	&	[0;1]	\\
Has secondary ability Flash	&	[0;1]	\\
Normalized main ability cooldown	&	[0;60]	\\
Normalized secondary ability cooldown	&	[0;60]	\\
Normalized health	&	[0;100]	\\
Normalized pitch	&	[-180;180]	\\
Normalized yaw	&	[0;360]	\\
Normalized reserve ammo	&	[0;48]	\\
Normalized magazine ammo	&	[0;12]	\\
\hline
\hline
\textbf{Bomb} &  \\
\hline
Has the bomb	&	[0;1]	\\
Teammate has the bomb	&	[0;1]	\\
Is dropping the bomb	&	[0;1]	\\
Is planting the bomb	&	[0;1]	\\
Is defusing the bomb	&	[0;1]	\\
Is bomb planted	&	[0;1]	\\
Normalized number of seconds of attempting to plant the bomb	&	[0;4]	\\
Normalized number of seconds of attempting to defuse the bomb	&	[0;7]	\\
Normalized number of seconds until explosion	&	[0;45]	\\
Normalized distance to bombsite	&	[0;10000]	\\
Normalized distance to bomb	&	[0;10000]	\\
Direction to bombsite (normalized x, y directional vector)	&	[0;10000]	\\
Direction to bomb (normalized x, y directional vector)	&	[0;10000]	\\
\hline
\hline
\textbf{Team} &  \\
\hline
Normalized time left in round	&	[0;120]	\\
Normalized min enemy distance	&	[0;10000]	\\
Normalized min enemy grenade distance	&	[0;10000]	\\
Normalized distance to teammate	&	[0;10000]	\\
Direction to teammate (normalized x, y directional vector)	&	[0;10000]	\\
\end{tabular}
\end{table}

\section{Human and Bot Gameplay Feature Distribution Comparisons}\label{sec:plots}
\begin{figure*} [!tb]
    \centering
    \includegraphics[width=0.9\textwidth]{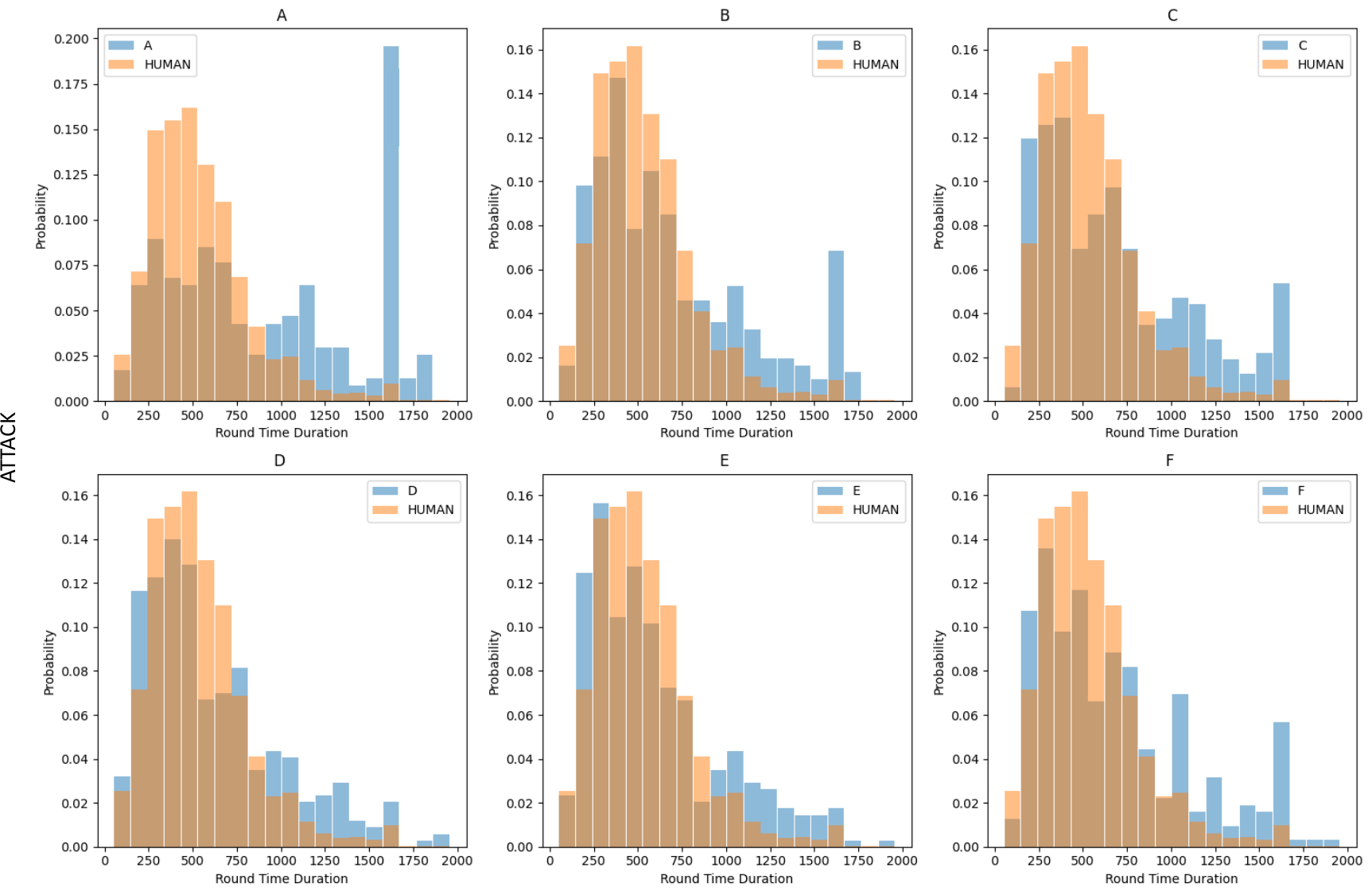}
    \caption{This figure shows the observed probabilities (y-axis) for observed  round durations (x-axis) while on the attacking team for the human play data (orange) and model play data (blue). Round Duration is shown in timesteps (16 values per second).}
    \label{fig:at_tim}
\end{figure*}

\begin{figure*} [!tb]
    \centering
    \includegraphics[width=0.9\textwidth]{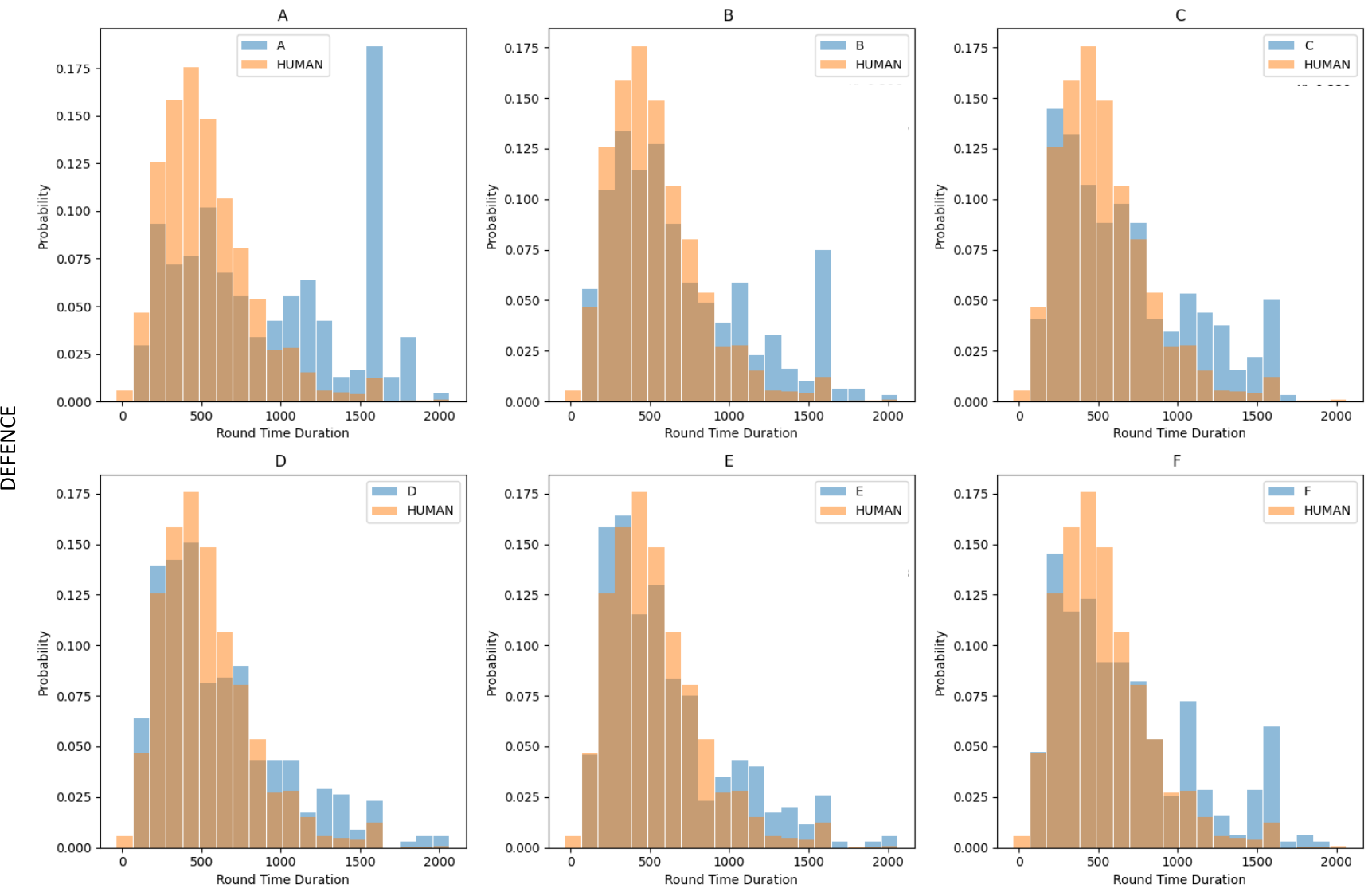}
    \caption{This figure shows the observed probabilities (y-axis) for observed  round durations (x-axis) while on the defending team for the human play data (orange) and model play data (blue). Round Duration is shown in timesteps (16 values per second).}
    \label{fig:df_tim}
\end{figure*}

\begin{figure*} [!tb]
    \centering
    \includegraphics[width=0.9\textwidth]{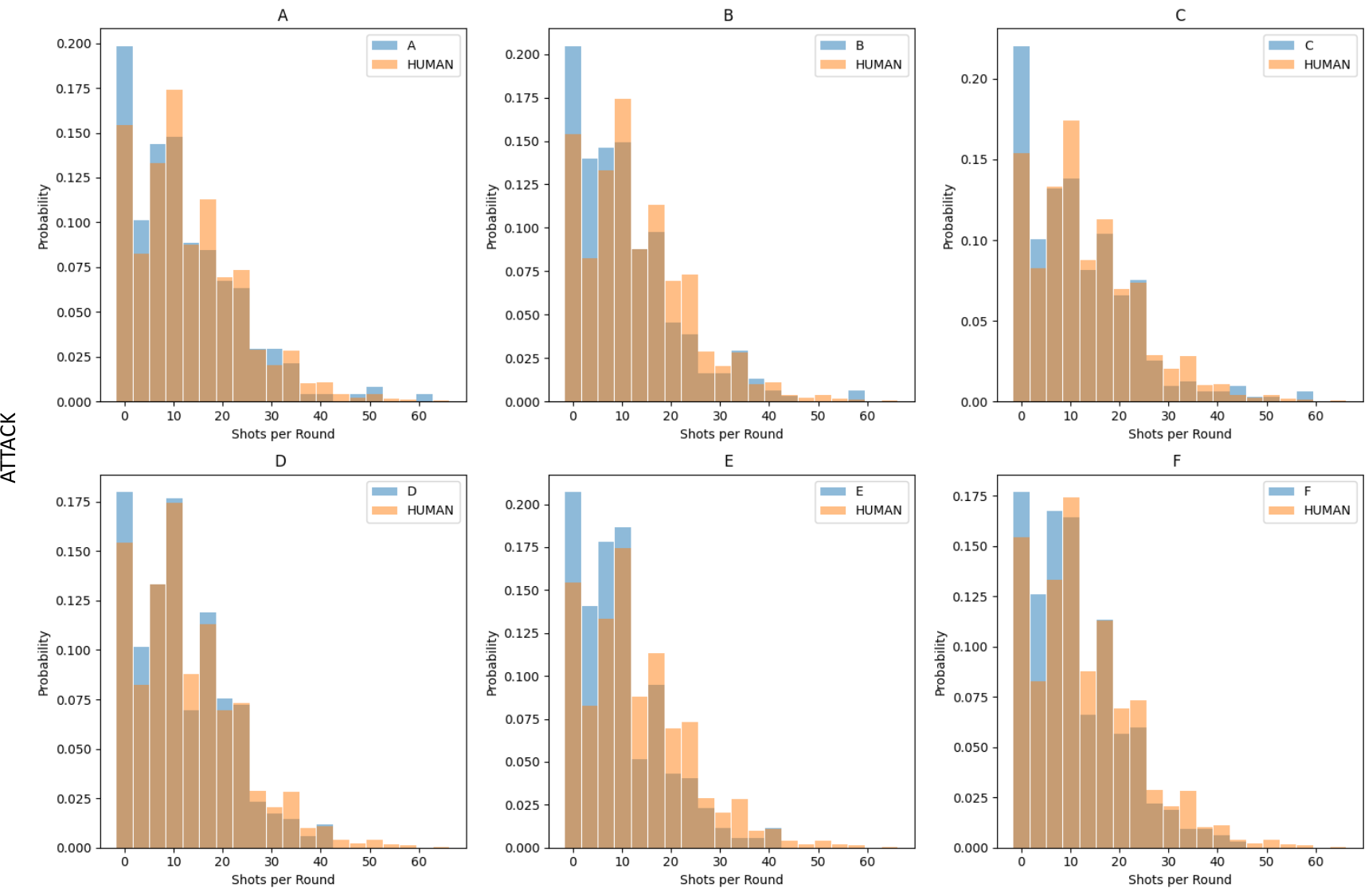}
    \caption{
    This figure shows the observed probabilities (y-axis) for the number of shots fired per round (x-axis) while on the defending team for the human play data (orange) and model play data (blue). }
    \label{fig:at_sh}
\end{figure*}

\begin{figure*} [!tb]
    \centering
    \includegraphics[width=0.9\textwidth]{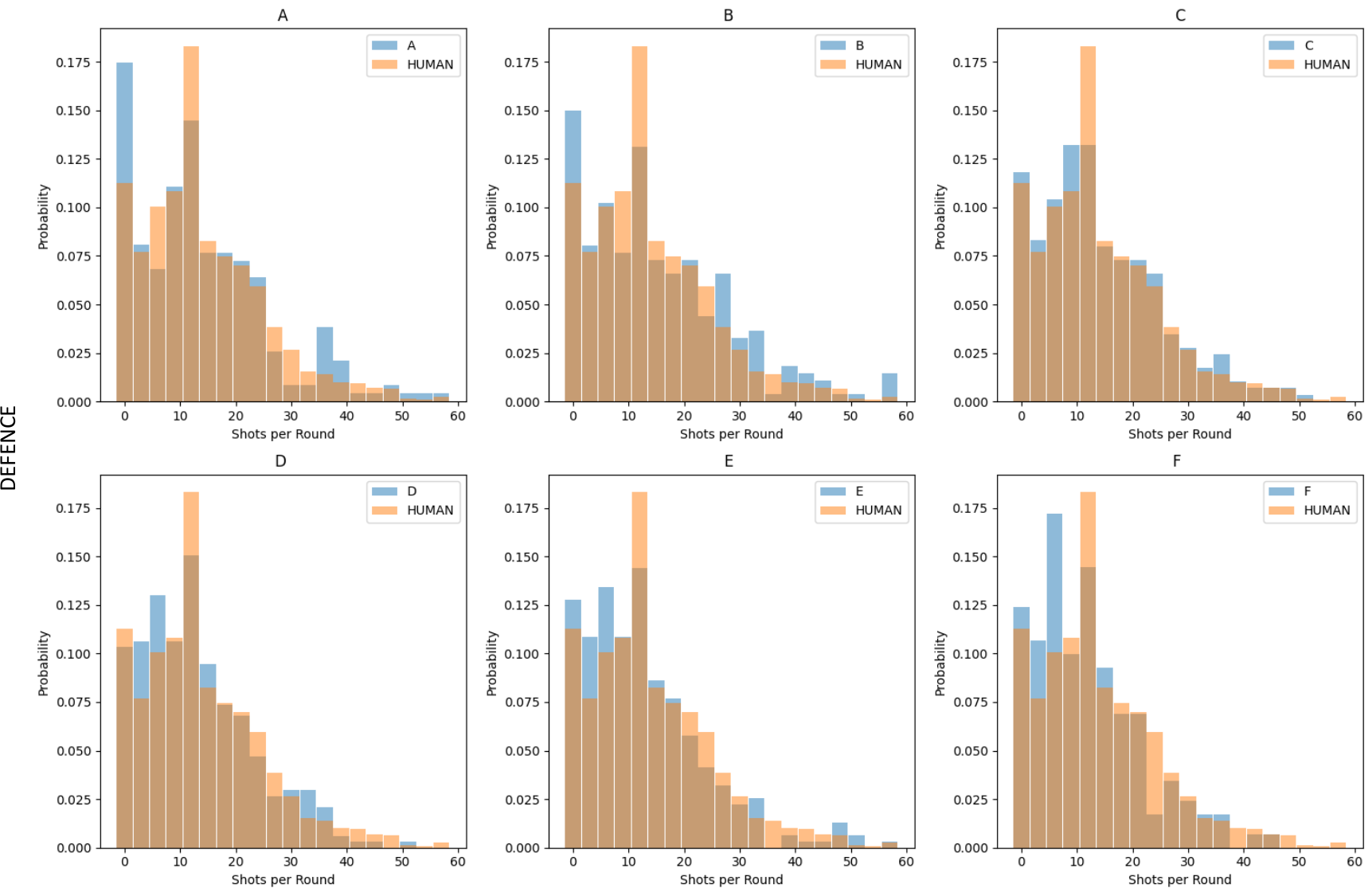}
    \caption{This figure shows the observed probabilities (y-axis) for the number of shots fired per round (x-axis) while on the attacking team for the human play data (orange) and model play data (blue). }
    \label{fig:df_sh}
\end{figure*}

\begin{figure*} [!tb]
    \centering
    \includegraphics[width=0.9\textwidth]{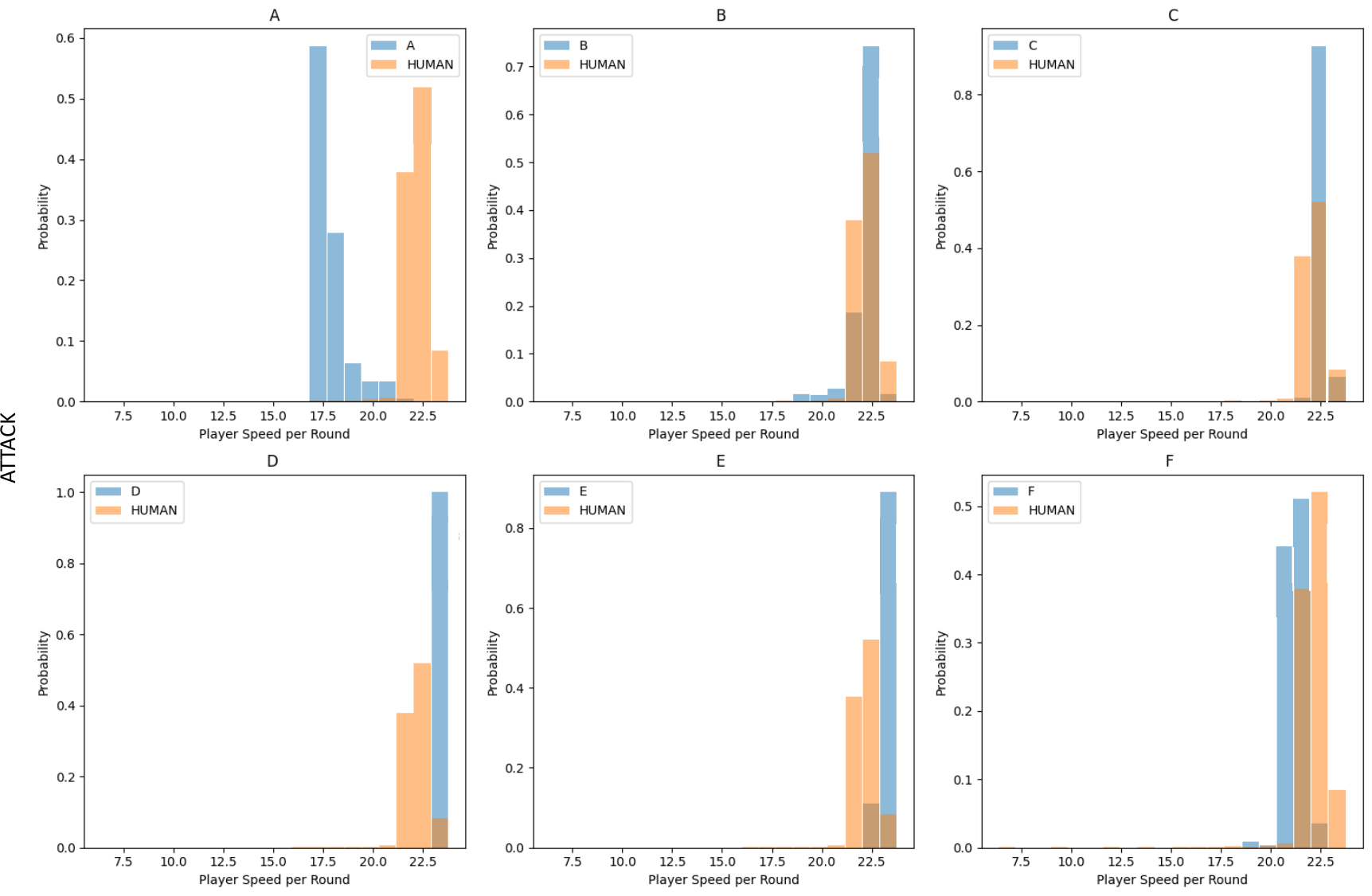}
    \caption{This figure shows the observed probabilities (y-axis) for the average movement speed (x-axis) while on the attacking team for the human play data (orange) and model play data (blue). }
    \label{fig:at_sp}
\end{figure*}

\begin{figure*} [!tb]
    \centering
    \includegraphics[width=0.9\textwidth]{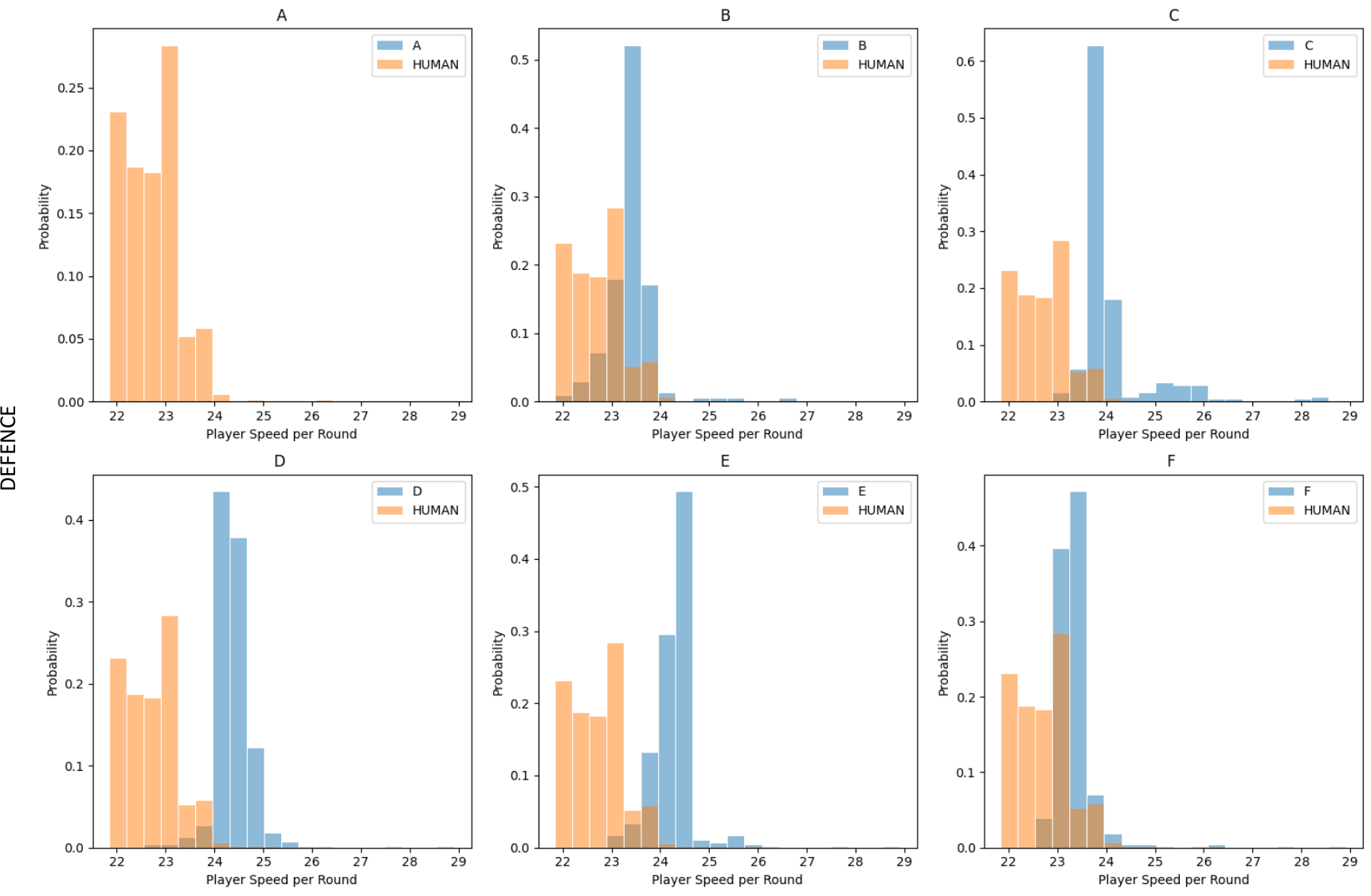}
    \caption{This figure shows the observed probabilities (y-axis) for the average movement speed (x-axis) while on the defending team for the human play data (orange) and model play data (blue). }
    \label{fig:df_sp}
\end{figure*}


\end{document}